\documentclass[11pt,nofootinbib,notitlepage,superscriptaddress]{revtex4-1}
\usepackage{amsmath,amssymb,float,color}
\usepackage[pdftex]{graphicx}

\def\bx{{\boldsymbol x}}

\def\bp{{\boldsymbol p}}

\def\bJ{{\boldsymbol J}}
\def\pr{\prime}
\def\pperp{p_{\!\perp}}
\def\bpperp{{\boldsymbol p}_{\!\perp}}
\def\bxperp{{\boldsymbol x}_{\!\perp}}
\def\tg{\text{g}}
\def\tq{\text{q}}

\def\slashchar#1{\setbox0=\hbox{$#1$}
\dimen0=\wd0 
\setbox1=\hbox{/} \dimen1=\wd1 
\ifdim\dimen0>\dimen1 
\rlap{\hbox to \dimen0{\hfil/\hfil}} 
#1 
\else 
\rlap{\hbox to \dimen1{\hfil$#1$\hfil}} 
/ 
\fi}

\begin{document}

\title{Nonequilibrium quark production in the expanding QCD plasma}

\author{Naoto Tanji}\email{ntanji@ectstar.eu}
\affiliation{Institut f\"{u}r Theoretische Physik, Universit\"{a}t Heidelberg, Philosophenweg 16, 69120 Heidelberg, Germany}
\affiliation{European Centre for Theoretical Studies in Nuclear Physics and Related Areas (ECT*) and Fondazione Bruno Kessler, Strada delle Tabarelle 286, I-38123 Villazzano (TN), Italy}

\author{J\"{u}rgen Berges}\email{berges@thphys.uni-heidelberg.de}
\affiliation{Institut f\"{u}r Theoretische Physik, Universit\"{a}t Heidelberg, Philosophenweg 16, 69120 Heidelberg, Germany}

\date{\today}

\begin{abstract}
We perform real-time lattice simulations of nonequilibrium quark production in the longitudinally expanding QCD plasma. 
Starting from a highly occupied gluonic state with vacuum quark sector, we extract the time evolution of quark and gluon number densities per unit transverse area and rapidity. The total quark number shows after an initial rapid increase an almost linear growth with time. Remarkably, this growth rate appears to be consistent with a simple kinetic theory estimate involving only two-to-two scattering processes in small-angle approximation. This extends previous findings about the role of two-to-two scatterings for purely gluonic dynamics in accordance with the early stages of the bottom-up thermalization scenario.
\end{abstract}

\maketitle

\section{Introduction and outline} \label{sec:intro}

Recently, much progress has been achieved in understanding the spatio-temporal evolution of the QCD plasma for relativistic heavy-ion collisions at sufficiently high energies. In such collisions a nonequilibrium plasma of highly occupied gluons with characteristic momentum $Q_s$ is expected to form shortly after the initial impact~\cite{Gelis:2010nm,Lappi:2006fp}. While the running QCD gauge coupling $\alpha_s(Q_s)$ is weak for sufficiently large $Q_s$, strong correlations occur because the typical gluon occupancy $f_g(Q_s) \sim 1/\alpha_s(Q_s)$ is high. As a consequence, perturbative descriptions are not expected to apply. However, in this regime the nonequilibrium quantum dynamics can be mapped onto a classical-statistical field theory description with quantum initial conditions, which can be solved with lattice simulation techniques.  
For instance, real-time lattice simulations identified which thermalization scenario is realized in the limit of very high collision energies~\cite{Berges:2013eia,Berges:2013fga}. The results have been implemented into effective kinetic descriptions to compute the later stages and thermalization of the quark-gluon plasma~\cite{Kurkela:2015qoa,Kurkela:2017hgm}. Subleading quantum corrections, including dynamical quarks, have been taken into account in fixed-box lattice simulations~\cite{Gelfand:2016prm}, imposing boost invariance for 2+1 dimensional simulations~\cite{Gelis:2005pb} and in kinetic descriptions with longitudinal expansion~\cite{Tanji:2017suk}.

To some extend the progress in our understanding of the complex many-body dynamics is based on the presence of attractors in the space-time evolution of the longitudinally expanding non-Abelian plasma~\cite{Berges:2013eia,Berges:2013fga,Tanji:2017suk}. Attractors lead to a certain insensitivity of the dynamics at later times to variations of the initial conditions. As a consequence, they diminish the problem of our insufficient knowledge about the systems's details at the time of collisional impact. Moreover, attractors imply another aspect of great practical importance: they reflect the presence of a large number of irrelevant operators in the renormalization group sense, such that rather simple effective descriptions of the relevant dynamics may be expected.\footnote{For work on nonthermal renormalization group fixed points in scalar/fermionic theories see Refs.~\cite{Berges:2008wm,Berges:2008sr,Berges:2010zv} and relations to gauge theories see Refs.~\cite{Berges:2014bba,Berges:2015ixa}; for recent work on hydrodynamic attractors see e.g.~Refs.~\cite{Strickland:2017kux,Spalinski:2017mel,Romatschke:2017acs}.} In fact, previous lattice simulation results of the longitudinally expanding pure gauge theory identified, after a short period of highly complex dynamics with gluon number changing processes, a number conserving regime that is consistently described by two-to-two scattering processes~\cite{Berges:2013eia,Berges:2013fga} in agreement with the earlier stages of the bottom-up thermalization scenario~\cite{Baier:2000sb}. No consistent perturbative power counting has yet been able to justify this relatively simple effective description, since e.g.~also plasma instabilities would play a major role according to state-of-the-art descriptions~\cite{Bodeker:2005nv,Kurkela:2011ub}.  

In this work we present the first results on 3+1 dimensional real-time lattice simulations of the longitudinally expanding non-Abelian plasma with dynamical quarks. More precisely, we employ SU(2) lattice gauge field theory with  $N_f$ flavors of degenerate quarks at weak coupling. We compute the plasma's time evolution starting from a highly occupied gluonic state with vacuum quarks and study its dependence on the quark mass and $\alpha_s N_f$. To describe nonequilibrium quark production and its interplay with the gluon sector, we extract quark and gluon distribution functions for their number densities per unit transverse area and rapidity. For small enough $\alpha_s N_f$ the gluons turn out to be little affected by the backreaction of the quarks, which only changes as $\alpha_s N_f$ is increased. In contrast, the high gluon occupancies always result in a strongly enhanced quark production. We find that the total quark number shows after an initial rapid increase an almost linear growth with time. Remarkably, this growth rate appears to be consistent with a simple kinetic theory estimate involving only two-to-two scattering processes in small-angle approximation.

The paper is organized as follows: In Sec.~\ref{sec:formulation} we describe the framework of real-time lattice simulations with dynamical quarks in the longitudinally expanding geometry. This includes a detailed discussion of the initial conditions for the nonequilibrium evolution and how observables are extracted from the lattice data. Sec.~\ref{sec:kinetic} reviews some relevant formulae from effective kinetic theory that are employed to analyze aspects of our simulation data. In Sec.~\ref{sec:results} we present our numerical simulation results for a wide range of quark masses and $N_f$. After the conclusions in Sec.~\ref{sec:conclusion}, we provide two appendices on details about the treatment of fermion doublers on a real-time lattice in Sec.~\ref{sec:doub} and a discussion on lattice cutoff dependencies in Sec.~\ref{sec:cutoff}.

\section{Real-time dynamics on the longitudinally expanding lattice} \label{sec:formulation}

We consider a nonequilibrium situation with strong initial gauge fields $A\sim Q_s /g$ and vacuum quarks for weak coupling $g\ll 1$ with $\alpha_s = g^2/(4\pi)$. By systematic power counting in $g$, to leading order the quantum dynamics is described by the corresponding classical-statistical gluonic field theory for given quantum initial correlations. The subleading order involves genuine quantum corrections to the dynamical evolution, which requires taking into account the quarks~\cite{Aarts:1998td,Kasper:2014uaa} that is described in the following.

In this section, we present the framework to compute the real-time dynamics of SU($N_c$) gauge fields and $N_f$-flavor quarks in a longitudinally expanding geometry. We employ the co-moving coordinates $x^\mu = (\tau ,\bxperp ,\eta )$, where $\tau=\sqrt{t^2 -z^2}$ is the proper time, $\bx_{\!\perp} =(x,y)$ are the transverse coordinates and $\eta=\text{atanh} (z/t)$ is the space-time rapidity. (Here natural units will be employed where the reduced Planck constant and the speed of light 
are set to one.) The metric in this coordinate system is $g_{\mu \nu} = \text{diag} (1,-1,-1,-\tau^2)$. 
The space coordinates $(\bxperp ,\eta)$ are discretized on an anisotropic lattice of size $N_\perp \times N_\perp \times N_\eta$ and spacings $a_\perp \times a_\perp \times a_\eta $ with periodic boundary condition. The proper time $\tau$ is treated as a continuum variable in this formulation.

\subsection{Equations of motion} \label{subsec:EOM}
Throughout this paper, we employ the Fock-Schwinger gauge in which the temporal component of the gauge potential vanishes, $A_\tau=0$. 
On the spatial lattice with a continuum time variable, the gauge fields are represented by link variables $U_i (x)$, $U_\eta(x)$ and electric fields $E^i (x)$, $E^\eta (x)$, where $i=1,2$ denotes the transverse directions. 
The equations of motion for the ($c$-number) gauge fields in the presence of quarks are\footnote{%
The transverse electric fields $E^i$ have canonical mass-dimension one and $E^i/\tau$ corresponds to the physical electric field, while the longitudinal electric field $E^\eta$ has mass-dimension two as usual electric fields.}   
\begin{gather}
\partial_\tau U_i (x) = ig \frac{a_i}{\tau} E^i (x) U_i (x) \, , \\
\partial_\tau U_\eta (x) = ig a_\eta \tau E^\eta (x) U_\eta (x) \, ,
\end{gather}
\begin{align}
\partial_\tau E^i (x) 
&= -\sum_{j\neq i} \frac{\tau}{g a_i a_j^2} \text{Im}\! \left[ U_{i,j} (x) +U_{i,-j} (x) \right]_\text{traceless} \notag \\
&\hspace{12pt} 
-\frac{1}{g \tau a_i a_\eta^2} \text{Im}\! \left[ U_{i,\eta} (x) +U_{i,-\eta} (x) \right]_\text{traceless}
-\tau J^i (x) \, , 
\label{eomET}
\end{align}
and
\begin{align}
\partial_\tau E^\eta (x) 
&= -\sum_{i=1,2} \frac{1}{g \tau a_\eta a_i^2} \text{Im}\! \left[ U_{\eta, i} (x) +U_{\eta,-i} (x) \right]_\text{traceless} 
-J^\eta (x) \, .
\label{eomEL}
\end{align}
Here, $U_{\mu ,\nu} (x)$ and $U_{\mu,-\nu} (x)$ are the plaquettes variables given by
\begin{equation}
U_{\mu ,\nu} (x) = U_\mu (x) U_\nu (x+\hat{\mu}) U_\mu^\dagger (x+\hat{\nu}) U_\nu^\dagger (x) \, ,
\end{equation}
and
\begin{equation}
U_{\mu,-\nu} (x) = U_\mu (x) U_\nu^\dagger (x+\hat{\mu}-\hat{\nu}) U_\mu^\dagger (x-\hat{\nu}) U_\nu (x-\hat{\nu}) \, .
\end{equation}
The subscript `traceless' means
\begin{equation}
\left[ X \right]_\text{traceless} = X -\frac{1}{N_c} \text{tr} (X) \, .
\end{equation}
In Eqs.~\eqref{eomET} and \eqref{eomEL}, $J^i(x)$ and $J^\eta (x)$ denote the color currents induced by the quark fields, which are responsible for the backreaction of quarks onto the gauge fields. Their explicit expressions are presented below [Eq.~\eqref{current0}]. Gauge field expectation values and correlation functions are obtained from sampling the initial configurations for the gauge fields (as described in Sec.~\ref{subsec:initial}) that are evolved according to the above equations of motion. 

To the order considered, the quark dynamics is governed by the Dirac equation in the presence of the sampled gauge fields. In the expanding geometry, the quark field operator $\hat{\psi} (x)$ obeys \cite{Gelis:2015eua}
\begin{equation}
\left[ i\gamma^0 \partial_\tau +i\gamma^i D_i +\frac{i}{\tau} \gamma^3 D_\eta -m\right] \hat{\psi} (x) = 0 \, ,
\label{Dirac0}
\end{equation}
where summation over the transverse index $i=1,2$ is implied. 
As a lattice covariant derivative, we employ the $\mathcal{O} (a^3)$-improved expression~\cite{Mueller:2016ven,Mace:2016shq}
\begin{align}
D_\mu \psi (x) 
&= \frac{c_1}{2a_\mu} \left[ U_\mu (x) \psi (x+\hat{\mu} ) -U_\mu^\dagger (x-\hat{\mu} ) \psi (x-\hat{\mu}) \right] \notag \\
&\hspace{10pt} 
+\frac{c_2}{2a_\mu} \left[ U_\mu (x) U_\mu (x+\hat{\mu}) \psi (x+2\hat{\mu} ) -U_\mu^\dagger (x-\hat{\mu} ) U_\mu^\dagger (x-2\hat{\mu} ) \psi (x-2\hat{\mu}) \right] \, ,
\end{align}
with coefficients $c_1=4/3$ and $c_2=-1/6$. 
Here we have written down the Dirac equation for `naive' fermions. Ways to remove fermion doublers are discussed in Sec.~\ref{subsec:initial} and Appendix \ref{sec:doub}. 
We have assumed, for simplicity, that all fermion flavors are degenerated. 
In terms of the quark field operator, the color current components appearing in Eqs.~(\ref{eomET}) and (\ref{eomEL}) read
\begin{align}
J^{a,\mu} (x) 
&= \frac{g}{2} N_f \, \text{Re} \big\langle \Omega_0 \big| \left( 
c_1\left[\, \overline{\hat{\psi}} (x) \, \raisebox{-3pt}{,}\ T^a \gamma^\mu U_\mu (x) \hat{\psi} (x+\hat{\mu}) \right] \right. \notag \\
&\hspace{10pt} 
+c_2\left[\, \overline{\hat{\psi}} (x) \, \raisebox{-3pt}{,}\ T^a \gamma^\mu U_\mu (x) U_\mu (x+\hat{\mu}) \hat{\psi} (x+2\hat{\mu}) \right]  \notag \\
&\hspace{10pt} \left. 
+c_2\left[\, \overline{\hat{\psi}} (x-\hat{\mu}) \, \raisebox{-3pt}{,}\ T^a \gamma^\mu U_\mu (x-\hat{\mu}) U_\mu (x) \hat{\psi} (x+\hat{\mu})  \right] \right) \big|\Omega_0 \big\rangle \, ,
\label{current0}
\end{align}
where $T^a$ $(a=1,\cdots ,N_c^2-1)$ are the generators of the gauge group SU($N_c$), and $\langle \Omega_0| \cdot |\Omega_0 \rangle$ denotes the expectation with respect to the initial quantum state.
For $\mu=\eta$, $\gamma^\mu$ is to be understood as $\gamma^3$. 

The color current is proportional to $gN_f$, while we consider strong initial color fields of the order of $1/g$. 
Therefore, the relative strength of the quark backreaction to the gauge fields is governed by the factor $g^2 N_f$ in this case.

\subsection{Stochastic low-cost method for quark dynamics} \label{subsec:stochastic}
The Dirac equation \eqref{Dirac0} is an operator equation, which in general is difficult to solve on a computer. Thanks to the fact that Eq.~\eqref{Dirac0} is linear with respect to the field operator, instead one may consider without loss of generality a $c$-number equation for the fermion mode functions, which can be solved numerically \cite{Aarts:1998td}. The mode functions $\psi_{\bpperp ,\nu ,s,c}^\pm (x)$ are introduced by the mode expansion of the field operator,
\begin{equation}
\hat{\psi} (x) = \sum_{s=\uparrow, \downarrow} \sum_{c=1}^{N_c} \frac{1}{L_\perp^2 L_\eta} \sum_{\bpperp ,\nu} \left[ \psi_{\bpperp ,\nu ,s,c}^+ (x) a_{\bpperp ,\nu ,s,c} +\psi_{\bpperp ,\nu ,s,c}^- (x) b_{\bpperp ,\nu ,s,c}^\dagger \right] \, ,
\end{equation}
where $a_{\bpperp ,\nu ,s,c}$ and $b_{\bpperp ,\nu ,s,c}$ are annihilation operators of a quark and an anti-quark, respectively, with spin $s$, color $c$ and momenta $(\bpperp ,\nu)$, which are conjugate to $(\bxperp ,\eta)$. 
The superscripts `$+$' and `$-$' for the mode functions distinguish positive and negative energy solutions. 
Flavor indices are omitted since we assume that all flavors are degenerate. 
The volume factor $L_\perp^2 L_\eta$ is comprised of the linear system sizes in the transverse directions $L_\perp =N_\perp a_\perp$ and in the longitudinal direction $L_\eta =N_\eta a_\eta$. 
The mode functions satisfy the same equation as the field operator,
\begin{equation}
\left[ i\gamma^0 \partial_\tau +i\gamma^i D_i +\frac{i}{\tau}\gamma^3  D_\eta -m\right] \psi_{\bpperp ,\nu ,s,c}^\pm (x) = 0 \, . 
\label{Dirac1}
\end{equation}
If one solves this equation with an appropriate initial condition, which will be specified in in Sec.~\ref{subsec:initial}, one can compute expectation values of any quark operator. 
For example, the expectation value of the commutator of two quark fields, i.e.~the fermion `statistical' two-point function, is defined by
\begin{equation}
F(x,y) = \frac{1}{2} \big\langle 0\big| \left[ \hat{\psi} (x) \, \raisebox{-3pt}{,}\ \overline{\hat{\psi}} (y) \right] \big| 0\big\rangle \, . 
\label{F0}
\end{equation}
It can be expressed in terms of the mode functions as
\begin{equation}
F(x,y) = \frac{1}{2} \sum_{s,c} \frac{1}{L_\perp^2 L_\eta} \sum_{\bpperp ,\nu} \left[ 
\psi_{\bpperp ,\nu ,s,c}^+ (x) \overline{\psi^+}_{\hspace{-5pt} \bpperp ,\nu ,s,c} (y)
-\psi_{\bpperp ,\nu ,s,c}^- (x) \overline{\psi^-}_{\hspace{-5pt} \bpperp ,\nu ,s,c} (y) \right] \, .
\label{F1}
\end{equation}

The numerical cost to solve the equations for the mode functions is proportional to $(N_c N_\perp^2 N_\eta)^2$, which is not amenable to large lattices. The numerical effort may be reduced by a stochastic method employing `low-cost' fermions~\cite{Borsanyi:2008eu}, where we will use a variant of this method~\cite{Gelis:2015eua,Gelis:2015kya}. In this approach, one employs two kinds of stochastic fermion fields corresponding to positive and negative energy modes,
\begin{equation}
\Psi^\pm (x) = \sum_{s,c} \frac{1}{L_\perp^2 L_\eta} \sum_{\bpperp ,\nu} \psi_{\bpperp ,\nu ,s,c}^\pm (x) c_{\bpperp ,\nu ,s,c}^\pm \, .
\label{stoPsi0}
\end{equation}
Here, $c_{\bpperp ,\nu ,s,c}^\pm$ are complex random Gaussian numbers whose ensemble averages satisfy
\begin{gather}
\langle c_{\bpperp ,\nu ,s,c}^\epsilon (c_{\bpperp^\pr ,\nu^\pr ,s^\pr,c^\pr}^{\epsilon^\pr})^* \rangle = L_\perp^2 L_\eta \, \delta^{\epsilon , \epsilon^\pr} \delta_{s,s^\pr} \delta_{c,c^\pr} \delta_{\bpperp , \bpperp^\pr } \delta_{\nu ,\nu^\pr} \, ,
\end{gather}
and
\begin{equation}
\langle c_{\bpperp ,\nu ,s,c}^\epsilon c_{\bpperp^\pr ,\nu^\pr ,s^\pr,c^\pr}^{\epsilon^\pr} \rangle = 0 \, . 
\end{equation}
The stochastic fermion fields obey the same Dirac equation
\begin{equation}
\left[ i\gamma^0 \partial_\tau +i\gamma^i D_i +\frac{i}{\tau} \gamma^3 D_\eta -m\right] \Psi^\pm (x) = 0 \, . 
\label{Dirac2}
\end{equation}
Once we compute the initial condition for them following Eq.~\eqref{stoPsi0}, the numerical cost to solve the Dirac equation for the stochastic fields is proportional to $N_c N_\perp^2 N_\eta N_\text{conf}$, where $N_\text{conf}$ is the number of configurations for the stochastic fields. The value of $N_\text{conf}$ depends on the quantity one computes and, of course, on the computational accuracy  demanded. In particular for space-averaged quantities, $N_\text{conf}$ can often be much smaller than $N_c N_\perp^2 N_\eta$, such that the stochastic fermion method is more efficient than the mode function method. 

In terms of the stochastic fields, the fermion statistical function \eqref{F0} can be computed by the ensemble average as
\begin{equation}
F(x,y) = \frac{1}{2} \sum_{\epsilon =\pm} \epsilon \, \langle \Psi^\epsilon (x) \overline{\Psi^\epsilon} (y) \rangle \, .
\label{F2}
\end{equation}
Similarly, the vacuum expectation of the color current operator is expressed by the fermion ensemble average as
\begin{align}
J^{a,\mu} (x) 
&= -\frac{g}{2} N_f \, \text{Re} \sum_{\epsilon =\pm} \epsilon \, \big\langle c_1\, \overline{\Psi^\epsilon} (x) T^a \gamma^\mu U_\mu (x) \Psi^\epsilon (x+\hat{\mu}) 
+c_2 \, \overline{\Psi^\epsilon} (x) T^a \gamma^\mu U_\mu (x) U_\mu (x+\hat{\mu}) \Psi^\epsilon (x+2\hat{\mu}) 
\notag \\
&\hspace{10pt}
+c_2 \, \overline{\Psi^\epsilon} (x-\hat{\mu}) T^a \gamma^\mu U_\mu (x-\hat{\mu}) U_\mu (x) \Psi^\epsilon (x+\hat{\mu})  \big\rangle \, .
\label{current}
\end{align}

\subsection{Initial conditions} \label{subsec:initial}
In the gauge sector, we initialize the fields according to a Gaussian initial density matrix~\cite{Berges:2013eia,Berges:2013fga}, which translates into 
\begin{align}
A_\mu^a (\tau_0 ,\bxperp ,\eta )
&= \sum_{\lambda=1,2} \frac{1}{L_\perp^2 L_\eta} \sum_{\bpperp ,\nu} \sqrt{f_\tg (\tau_0 ,\bpperp ,\nu )}
\left[ \xi_{\mu , \bpperp , \nu}^{(\lambda)} (\tau_0 ) e^{i\bpperp \cdot \bxperp +i\nu \eta} c_{\bpperp ,\nu}^{\lambda ,a} +\text{c.c} \right] , \label{iniA} \\
E^{a,\mu} (\tau_0 ,\bxperp ,\eta )
&= -\tau_0 g^{\mu \rho} \sum_{\lambda=1,2} \frac{1}{L_\perp^2 L_\eta} \sum_{\bpperp ,\nu} \sqrt{f_\tg (\tau_0 ,\bpperp ,\nu )}
\left[ \dot{\xi}_{\rho , \bpperp , \nu}^{(\lambda)} (\tau_0 ) e^{i\bpperp \cdot \bxperp +i\nu \eta} c_{\bpperp ,\nu}^{\lambda ,a} +\text{c.c} \right] , \label{iniE}
\end{align}
with Gaussian random numbers $c_{\bpperp ,\nu}^{\lambda ,a}$ that satisfy
\begin{equation}
\langle c_{\bpperp ,\nu}^{\lambda ,a} (c_{\bpperp^\pr ,\nu^\pr}^{\lambda^\pr ,a^\pr})^* \rangle
= L_\perp^2 L_\eta \, \delta^{\lambda ,\lambda^\pr} \delta^{a,a^\pr} \delta_{\bpperp ,\bpperp^\pr} \delta_{\nu ,\nu^\pr} 
\end{equation}
and
\begin{equation}
\langle c_{\bpperp ,\nu}^{\lambda ,a} c_{\bpperp^\pr ,\nu^\pr}^{\lambda^\pr ,a^\pr} \rangle = 0 \, .
\end{equation}
Here, $\xi_{\mu , \bpperp , \nu}^{(\lambda)} (\tau_0 ) $ are the transverse polarization vectors with polarization $\lambda$, and $\dot{\xi}_{\mu , \bpperp , \nu}^{(\lambda)} (\tau) =\partial_\tau \xi_{\mu , \bpperp , \nu}^{(\lambda)} (\tau)$.\footnote{The polarization vectors are constructed such that the Coulomb-type gauge condition
$\sum_i \partial_i A_i (x) +\tau^{-2} \partial_\eta A_\eta (x) = 0$ is satisfied at the initial time.} 
Their explicit forms can be found in Ref.~\cite{Berges:2013fga}. 
The factor $f_\tg (\tau_0 ,\bpperp ,\nu )$ corresponds to the initial gluon momentum distribution function. To describe the highly occupied plasma, we employ
\begin{equation}
f_\tg (\tau_0 ,\bpperp ,\nu ) = \frac{n_0}{g^2} \, \Theta \left(Q_s -\sqrt{\bpperp^2 +(\xi_0 \nu/\tau_0)^2}\right) \, ,
\label{inifg}
\end{equation}
where $Q_s$ is the characteristic (saturation) momentum  scale, $n_0$ parametrizes the initial overoccupation, and $\xi_0$ is the initial anisotropy parameter. 
The link variable $U_\mu$ and the gauge field \eqref{iniA} are related by
\begin{equation}
U_\mu (x) = \exp \left[ ig a_\mu A_\mu^a (x) T^a \right] \, .
\label{link}
\end{equation}
In this construction of the gauge fields, the Gauss law
\begin{equation}
\sum_{\mu =1,2,\eta} \frac{1}{a_\mu} \left[ E^\mu (x) -U_\mu^\dagger (x-\hat{\mu}) E^\mu (x-\hat{\mu}) U_\mu (x-\hat{\mu}) \right] = 0 
\end{equation}
is initially not satisfied. We restore the Gauss law at the initial time by using the relaxation method described in Ref.~\cite{Moore:1996qs}. If the Gauss law is satisfied at the beginning, the evolution equations preserve it within the accuracy of the time discretization employed.\footnote{We have used the fourth-order Runge-Kutta method for the time evolution.}

The overoccupied gluonic plasma that can be characterized by the distribution \eqref{inifg} is expected to appear after the coherent initial gauge fields decay by instabilities, and its typical time scale is parametrically $\tau_0 \sim Q_s^{-1} \ln^2 \alpha_s^{-1}$ \cite{Romatschke:2006nk}. 
In this study, we take the value of the coupling $g=10^{-2}$ and correspondingly we adopt the initial time $Q_s \tau_0=100$.

In the quark sector, we assume vacuum initial condition; all the expectations are computed with the vacuum state $|0\rangle$, and the quark mode functions are initialized to be free ones,
\begin{equation}
\psi_{\bpperp ,\nu ,s,c}^\pm (\tau_0 ,\bxperp ,\eta ) = \psi_{\bpperp ,\nu ,s,c}^{\text{free}\, \pm} (\tau_0 ,\bxperp ,\eta ) \, .
\label{inipsi}
\end{equation}
The expressions for the free spinors in the $\tau$-$\eta$ coordinates can be found in Ref.~\cite{Gelis:2015eua}. 
Our assumption corresponds to neglecting the quark production at earlier times before $\tau_0 \sim Q_s^{-1} \ln^2 \alpha_s^{-1}$. The quark production can happen at the instant of a collision ($\tau=0$) and also in the earlier stage of the Glasma evolution ($0<\tau \alt Q_s^{-1} \ln^2 \alpha_s^{-1}$)~\cite{Gelis:2005pb,Gelis:2015eua}, which can in principle be computed by the framework presented in this paper. However, real-time lattice computations in the expanding geometry at early times $\tau \alt Q_s^{-1}$ are extremely demanding because of the rapid change of longitudinal scales.
In the present study, we focus on the quark dynamics in the incoherent gluon plasma for times $Q_s \tau \ge 100$. 

\begin{figure}[tb]
 \begin{center}
  \includegraphics[clip,width=8cm]{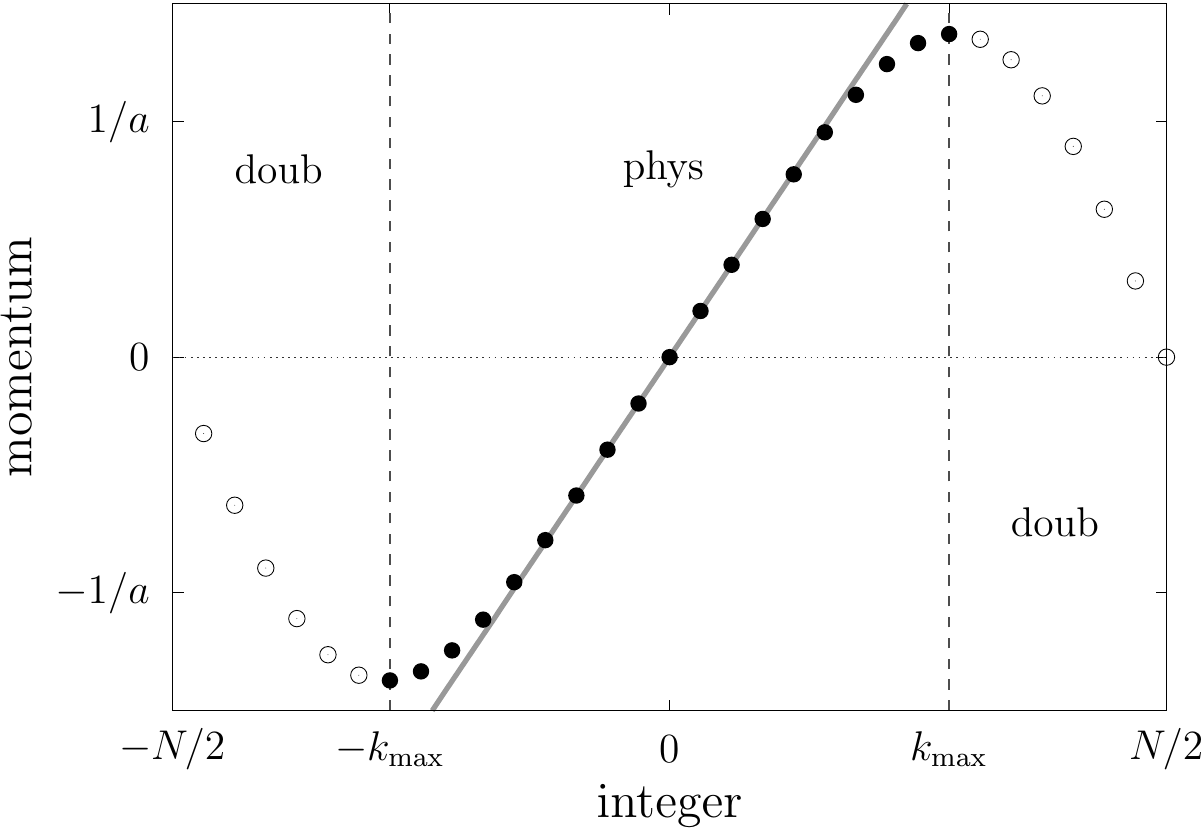} 
  \caption{A schematic plot of the lattice momentum \eqref{lattmom} as a function of the integer $k$. We call the modes in $-k_\text{max} \leq k \leq k_\text{max}$ physical modes, which are denoted by black filled circles, while the modes denoted by open circles correspond to doublers. The continuum dispersion is shown by a gray solid line.}
  \label{fig:lattmom}
 \end{center}
\end{figure}

The momenta $(p_x ,p_y ,\nu)$ in Eq.~\eqref{inipsi} are  lattice momenta that are related to integers by
\begin{equation}
p_x = \frac{c_1}{a_\perp} \sin \left( 2\pi \frac{k_x}{N_\perp} \right) +\frac{c_2}{a_\perp} \sin \left( 4\pi \frac{k_x}{N_\perp} \right) 
\label{lattmom}
\end{equation}
for $k_x=-N_\perp/2+1, \cdots ,0,\cdots ,N_\perp/2$, and similarly for the $y$ and $\eta$ components.
A schematic plot of the lattice momentum is shown in Fig.~\ref{fig:lattmom}. There are two zero-modes in each dimension (fermion doubling). We will refer to modes that exist between the maximum and the minimum of the lattice momentum as physical modes $\Lambda_\text{phys}$, which are denoted by black filled circles in Fig.~\ref{fig:lattmom}, while other modes are called doubler modes. Because of the improvement of the classical action, the regions of the physical modes and the doubler modes become asymmetric. The maximum of the lattice momentum occurs at
\begin{equation}
k_\text{max} = \bigg\lfloor \frac{N_\perp}{\pi} \text{atan} \sqrt{\frac{3+2\sqrt{6}}{5}} \bigg\rfloor 
\approx \lfloor 0.286 N_\perp \rfloor \, ,
\end{equation} 
where $\lfloor \ \rfloor$ denotes the floor function giving the largest integer that is less than or equal to its argument. 

We eliminate the doubler modes from the initial condition by initializing the fermion stochastic field as
\begin{equation}
\Psi^\pm (x) = \sum_{s,c} \frac{1}{L_\perp^2 L_\eta} \sum_{(k_x ,k_y ,k_\eta) \in \Lambda_\text{phys}} \psi_{\bpperp ,\nu ,s,c}^\pm (x) c_{\bpperp ,\nu ,s,c}^\pm
\label{stoPsi}
\end{equation}
instead of Eq.~\eqref{stoPsi0}.
On the other hand, we do not modify the evolution equation \eqref{Dirac2}. Therefore, the doubler modes can be, in principle, excited during the time evolution. However, as long as the momentum cutoffs are sufficiently large and we do not continue the time evolution for too long times, the contamination from doubler modes is found to be negligible \cite{Tanji:2015ata}.
We have explicitly checked this for our results as is described in more detail in Appendix~\ref{sec:doub}, where we also make comparisons to employing a generalized Wilson term for the suppression of doublers in the expanding geometry.\footnote{Employing a Wilson term, one needs to compute additional contributions to the evolution equations. Furthermore, one has to have smaller time steps compared to the naive fermion case in order to correctly compute the evolution of heavy doubler modes. As a consequence, suppressing doublers by initial conditions turns out to be very efficient for our purposes.}

\subsection{Observables} \label{subsec:obs}
As outputs of the lattice numerical computations, we 
compute correlation functions of gauge and quark fields. 
In order to be able to make comparisons also to effective kinetic descriptions, we mainly focus here on momentum distribution functions (occupation numbers) of gluons and quarks. 
The quasi-particle distribution function provides us with useful insights into the nonequilibrium evolution of the system, although it has shortcomings that it is gauge-dependent and no unique definition of quasi-particle numbers exist in the interacting theory. 

The quasi-particle distributions can be extracted from the equal-time two-point correlation functions \cite{Berges:2004yj}.
For the gluon distribution function, we employ the definition used in Ref.~\cite{Berges:2013eia,Berges:2013fga}:
\begin{align}
f_\tg (\tau ,\bpperp ,p_z ) 
&= \frac{\tau^2}{\nu_\tg L_\perp^2 L_\eta} \sum_{a=1}^{N_c^2-1} \sum_{\lambda=1,2}
\bigg\langle \bigg| a_\perp^2 a_\eta \sum_{\bxperp ,\eta} g^{\mu \rho} \left[ \left( \xi_{\mu, \bpperp ,\nu}^{(\lambda)} (\tau)\right)^* \overleftrightarrow{\partial_\tau} A_\rho^a (\tau ,\bxperp ,\eta ) e^{-i\bpperp \cdot \bxperp -i\nu \eta} \right] \bigg|^2 \bigg\rangle \, ,
\label{def_fg}
\end{align}
where $\nu_\tg=2(N_c^2-1)$ denotes the number of internal degrees of freedom. 
The gauge field $A_\mu^a (\tau ,\bxperp ,\eta )$ can be extracted from the link variable by the inverse of \eqref{link}, and the time derivative of the gauge field is related to the electric field as $E^{a,\mu} = \tau g^{\mu \nu} \partial_\tau A_\nu^a$. 
In the longitudinally expanding geometry, we take $p_z$ and $\nu$ as related by $p_z =\nu/\tau$. 
The bracket $\langle \ \rangle$ in Eq.~\eqref{def_fg} denotes the ensemble average over the random initial conditions for the gauge field, such that the residual gauge freedom in the Fock-Schwinger gauge ($A_\tau=0$) is fixed by the Coulomb-type gauge condition
\begin{equation}
\sum_i \partial_i A_i (x) +\tau^{-2} \partial_\eta A_\eta (x) = 0 \, .
\label{Coulomb}
\end{equation}
Although this condition is satisfied by the initial condition \eqref{iniA}, it is not preserved by the time evolution. When we compute the distribution functions for $\tau > \tau_0$, we transform the gauge field so that the condition \eqref{Coulomb} is fulfilled. Of course, the quark fields must be transformed accordingly. 

We define the quark distribution function in a similar way by the projection onto the free mode functions.
In terms of the stochastic fermion fields,
\begin{align}
f_\tq (\tau ,\bpperp ,p_z ) 
= \frac{1}{\nu_\tq L_\perp^2 L_\eta} \sum_{s=\uparrow, \downarrow} \sum_{c=1}^{N_c} &\bigg\langle \!\!\! \bigg\langle \bigg| a_\perp^2 a_\eta \sum_{\bxperp ,\eta} \left[ \left( \psi_{\bpperp ,\nu ,s,c}^{\text{free} \, +} (\tau ,\bxperp ,\eta ) \right)^\dagger \Psi^- (\tau ,\bxperp ,\eta ) \right] \bigg|^2 \notag \\
&\hspace{-2pt}
 +\bigg| a_\perp^2 a_\eta \sum_{\bxperp ,\eta} \left[ \left( \psi_{\bpperp ,\nu ,s,c}^{\text{free} \, -} (\tau ,\bxperp ,\eta ) \right)^\dagger \Psi^+ (\tau ,\bxperp ,\eta ) \right] \bigg|^2 \bigg\rangle \!\!\! \bigg\rangle \, ,
\label{def_fq}
\end{align}
where $\nu_\tq = 4 N_c N_f$ is the number of the internal degrees of freedom for quarks and anti-quarks. As noted before, the flavor indexes are omitted.
The double bracket $\langle \!\!\! \langle \ \rangle \!\!\! \rangle$ denotes the ensemble averages over both the initial gauge field configurations and the stochastic fermion field configurations.
This definition corresponds to the expectation of the number operators averaged over all degrees of freedom,
\begin{equation}
f_\tq (\tau ,\bpperp ,p_z ) 
= \frac{1}{\nu_\tq} \sum_{s=\uparrow, \downarrow} \sum_{c=1}^{N_c} \langle 0| \left( a_{\bpperp ,\nu ,s,c}^\dagger a_{\bpperp ,\nu ,s,c} +b_{\bpperp ,\nu ,s,c}^\dagger b_{\bpperp ,\nu ,s,c} \right) |0\rangle \, .
\end{equation}
If applied to the non-expanding system, our definition is equivalent to the definition in terms of the statistical two-point function employed in Refs.~\cite{Berges:2010zv,Berges:2013oba,Gelfand:2016prm}.
As with the gluon distribution, we compute the quark distribution after doing the gauge transformation to the Coulomb gauge \eqref{Coulomb}. 

We will also frequently consider the total quasi-particle number densities per unit transverse area and per unit rapidity:
\begin{equation}
\frac{dN_\text{g/q}}{d^2 x_{\!\perp} d\eta} 
= \frac{\nu_\text{g/q}}{L_\perp^2 L_\eta} \sum_{\bpperp ,\nu} f_\text{g/q} (\tau ,\bpperp ,p_z =\nu/\tau) \, . 
\label{def_num}
\end{equation} 
For quarks, this quantity involves the total particle number of quarks and anti-quarks.

\section{Effective kinetic theory diagnostics} \label{sec:kinetic}

To gain physical insight into results from lattice simulations, we will make comparisons with effective kinetic theory for QCD at weak coupling as described in Ref.~\cite{Arnold:2002zm}, and applied to numerical computations in the expanding geometry in Refs.~\cite{Kurkela:2015qoa,Keegan:2015avk,Keegan:2016cpi}. To this end, we give in this section some relevant formulae that will be employed as diagnostic tools to help analyzing the lattice simulation data.

We emphasize that since we are considering the over-occupied regime with very high gluon occupancies $\sim 1/\alpha_s$ it cannot be approximated by power counting in $\alpha_s$ underlying kinetic descriptions. What is remarkable is the fact that simple/naive estimates seem to give nevertheless the right order of magnitude of quark production from over-occupied gluons at least for integrated quantities, which we will be focusing on. 
  
The kinetic equations for gluons and quarks in the expanding system are of the form
\begin{gather}
\left( \frac{\partial}{\partial \tau} -\frac{p_z}{\tau} \frac{\partial}{\partial p_z} \right) f_\tg (\tau ,\bp )
= C_\tg [f_\tg ,f_\tq] \, , \\
\left( \frac{\partial}{\partial \tau} -\frac{p_z}{\tau} \frac{\partial}{\partial p_z} \right) f_\tq (\tau ,\bp )
= C_\tq [f_\tq ,f_\tg] \, ,
\end{gather}
where $C_\text{g/q}$ denote the collision terms. 
In the effective kinetic theory to leading order the collision terms involve $2\leftrightarrow 2$ scattering processes as well as effective $1\leftrightarrow 2$ processes. However, earlier lattice simulation results for the pure gauge theory indicate that the earlier-stage dynamics of the plasma in weak coupling may be characterized by $2\leftrightarrow 2$ scatterings~\cite{Berges:2013eia,Berges:2013fga} in accordance with the bottom-up thermalization scenario \cite{Baier:2000sb}. 
Therefore, we will compare our lattice results in the time regime considered to a kinetic description that takes into account only $2\leftrightarrow 2$ scattering processes. 

In the comparison to lattice results, we will deal with the integrated total number densities defined in Eq.~\eqref{def_num}. 
By integrating the kinetic equations over spatial momenta, one can relate the time derivative of the number density to the collision terms as
\begin{equation}
\frac{dN_\text{g/q}}{d\tau d^2 x_{\!\perp} d\eta} 
= \frac{\nu_\text{g/q}}{L_\perp^2 L_\eta} \sum_{\bpperp ,\nu} C_\text{g/q} \, .
\label{dndt0}
\end{equation}
To simplify the analysis, we furthermore apply the small-angle approximation, which is expected to be justified for long-ranged processes such that an exchanged particle is massless \cite{lifshitz1981physical,Mueller:1999pi}. 
By the small-angle approximation, each collision term takes the form of the sum of a diffusion term and a source term $S_\text{g/q}$~\cite{Blaizot:2014jna},
\begin{gather}
C_\text{g/q} = -\nabla_\bp \cdot \bJ_\text{g/q} +S_\text{g/q} \, .
\end{gather}
The diffusion terms do not contribute to the momentum integral in Eq.~\eqref{dndt0}. 
The source terms read
\begin{gather}
S_\tg (\tau ,\bp ) = \frac{g^4}{4\pi} C_F N_f \mathcal{L} I_c \, \frac{1}{p} 
\left[ f_\tq (\tau ,\bp ) \left(1+f_\tg (\tau ,\bp )\right) -f_\tg (\tau ,\bp ) \left(1-f_\tq (\tau ,\bp )\right) \right] \, , \label{Sg} \\
S_\tq (\tau ,\bp ) = -\frac{g^4}{4\pi} C_F^2 \mathcal{L} I_c \, \frac{1}{p} 
\left[ f_\tq (\tau ,\bp ) \left(1+f_\tg (\tau ,\bp )\right) -f_\tg (\tau ,\bp ) \left(1-f_\tq (\tau ,\bp )\right) \right] \, , \label{Sq}
\end{gather}
where $p=\sqrt{\bpperp^2 +p_z^2}$, $C_F=(N_c^2-1)/(2N_c)$ and
\begin{equation}
I_c (\tau ) = \frac{1}{\tau L_\perp^2 L_\eta} \sum_{\bpperp ,\nu} \frac{1}{p} \left[ f_\tg (\tau ,\bp ) +f_\tq (\tau ,\bp )\right] \, .
\end{equation}
Here, $\mathcal{L}$ denotes the `Coulomb logarithm', which encodes the infrared divergence of the scattering amplitude regulated by the Debye mass scale, 
\begin{equation}
\mathcal{L} = \int_{q_\text{min}}^{q_\text{max}} \frac{dq}{q} \, .
\end{equation}
As cutoffs, we employ $q_\text{max}=Q_s$ and $q_\text{min}=m_D$, where $m_D$ is the Debye mass scale given in terms of the distribution functions as
\begin{equation}
m_D^2 = 4g^2 \frac{1}{\tau L_\perp^2 L_\eta} \sum_{\bpperp ,\nu} \frac{1}{p} \left[ N_c f_\tg (\tau ,\bp ) +N_f f_\tq (\tau ,\bp )\right] \, .
\label{Debye}
\end{equation}
This expression denotes the screening mass for gluons, while that for quarks has different numerical factors of order one in front of $f_\text{g}$ and $f_\text{q}$ \cite{Arnold:2002zm}. 
When an exchanged particle is a quark, it is more adequate to use the screening mass for quarks as the infrared cutoff. However, here we only want to set the characteristic scale. Moreover,
the difference between the two screening masses may be absorbed by the ambiguity in the choices of the cutoff scales of the Coulomb logarithm in the small-angle approximation for the kinetic equation, which results in a logarithmic uncertainty for the production rate. 

Since we consider only 2$\leftrightarrow$2 scatterings, the total particle number is conserved, 
which is guaranteed by the relation $\nu_\tg S_\tg +\nu_\tq S_\tq = 0$. 

By substituting Eq.~\eqref{Sq} into Eq.~\eqref{dndt0}, we obtain
\begin{equation}
\frac{dN_\tq}{d\tau d^2 x_{\!\perp} d\eta} 
=  \frac{g^4}{4\pi} \frac{(N_c^2-1)^2}{N_c} N_f \, \mathcal{L} I_c \, \frac{1}{L_\perp^2 L_\eta} \sum_{\bpperp ,\nu} \frac{1}{p} 
\left[ f_\tg (\tau,\bp ) \left(1-2f_\tq (\tau,\bp )\right) -f_\tq (\tau,\bp ) \right] \, . \label{kine_nq}
\end{equation}
The process contributing to this production rate is $\tg \tg \leftrightarrow \tq \bar{\tq}$. 
The factor $(1-2f_\tq)$ reflects the phenomenon of Pauli blocking for fermionic degrees of freedom. In the overoccupied gluon plasma with $f_\tg \gg 1$, the last term in the square bracket is negligible, while the Pauli blocking term is not negligible as is discussed in Sec.~\ref{sec:results}. 

The rather simple expressions above will be employed to further analyze lattice simulation results in the following way. We will evaluate the effective kinetic theory expressions by inserting on their right hand sides the distribution functions obtained from lattice simulation data. The left-hand-side outcomes of these expressions can be compared with results directly extracted from lattice simulations according to Eq.~\eqref{def_num}. Comparing both gives information about the validity of the simplifying assumptions underlying the effective kinetic description employed. Such an agreement provides a necessary condition for the successful description of the lattice QCD dynamics by the effective kinetic theory. A comparison of the lattice data to a separate dynamical solution of the full kinetic theory with quarks is beyond the scope of the present work, and our analysis in this respect should be thought of as providing evidence that such an enterprise is worth doing.

\section{Lattice simulation results and comparison to kinetic estimates} \label{sec:results}

In this section we present numerical results for the lattice QCD simulations with $N_c=2$ in the expanding geometry.  
The values of the coupling constant, the initial time, and the initial anisotropy parameter are fixed to $g=10^{-2}$, $Q_s \tau_0=100$, and $\xi_0=2$.
Quarks with $N_f$-flavors are taken to have a common mass $m$. At first, we consider $N_f=1$. In this case, the backreaction from the quark sector to the gluon sector is expected to be negligible since its strength is governed by the factor $g^2 N_f$ according to Sec.~\ref{subsec:EOM}.
In Sec.~\ref{subsec:largeNf}, we consider the case of larger $N_f$ such that the effects of the backreaction may not be neglected. Unless stated otherwise, the lattice parameters used in computations for this section are $N_\perp=48$, $N_\eta=256$, $Q_s a_\perp =0.625$, $a_\eta=1.95\times 10^{-3}$. We discuss the dependence on the lattice parameters in Appendix~\ref{sec:cutoff}. 

\subsection{Time evolution} \label{subsec:time}
We first investigate the time evolution of the distribution functions and the number densities for $N_f=1$ and $m/Q_s=0.1$. 

\begin{figure}[tb]
 \begin{center}
  \includegraphics[clip,width=8cm]{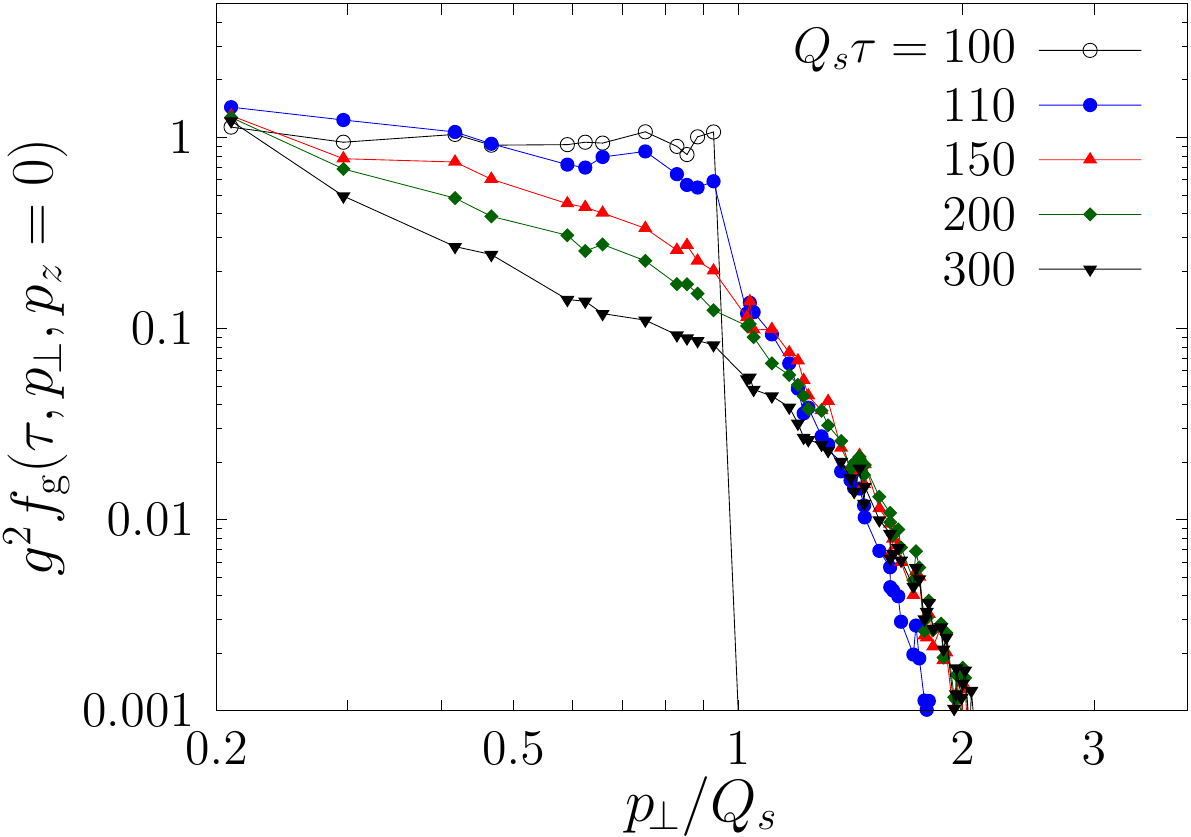} 
  \caption{Time evolution of the gluon transverse momentum distribution evaluated at $p_z=0$ for different times $Q_s \tau$.}
  \label{fig:distri_gpt}
 \end{center}
\end{figure}

In Fig.~\ref{fig:distri_gpt}, we show the gluon distribution $f_\tg$ as a function of transverse momentum $\pperp =\sqrt{p_x^2+p_y^2}$ evaluated at $p_z=0$ at different times $Q_s \tau$. Since the initial gluon occupation number is as large as $n_0/g^2$, we plot $g^2 f_\tg$. The initial occupation parameter is set to $n_0=1$. Due to infrared and ultraviolet cascades, the occupation numbers in the low and high momentum regions grow at early times, while the intermediate momentum region decreases. In fact, the results for the gluon sector are in line with previous studies in the absence of backreactions from the quarks as expected from the smallness of the factor $g^2 N_f = 10^{-4}$ employed. The gluon distribution approaches a nonthermal fixed point characterized by a self-similar scaling behavior \cite{Berges:2013eia,Berges:2013fga} reflecting, in particular, the longitudinal expansion and the longitudinal momentum broadening due to scatterings. To precisely compute that later-time behavior, one needs to employ larger lattices than those employed in the present study. Since the numerical cost to simulate fermion fields is much higher than the case of pure Yang-Mills simulations, we restrict ourself to the time range $Q_s \alt 300$, in which lattice artifacts can be well controlled with the lattice sizes we employ in this study. For more detailed discussions on the pure gauge theory dynamics, we refer readers to Refs.~\cite{Berges:2013eia,Berges:2013fga}, and we will mainly focus on the quark sector when backreactions onto the gluons play no important role.

\begin{figure}[tb]
 \begin{center}
  \includegraphics[clip,width=8cm]{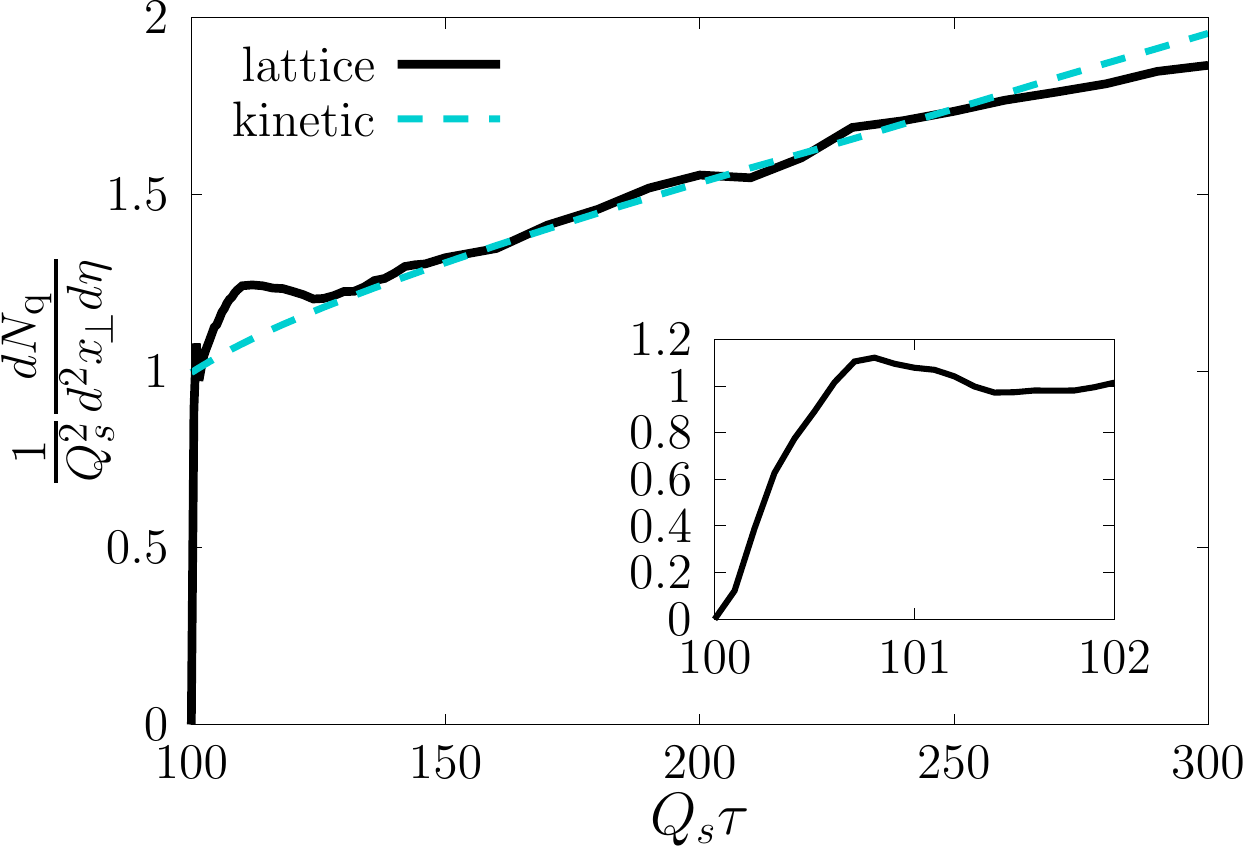} 
  \caption{Time evolution of the quark number density. The inset shows the early-time behavior.
The lattice result is compared to the kinetic theory estimate based on $2\leftrightarrow 2$ scatterings.}
  \label{fig:numq}
 \end{center}
\end{figure}

In order to analyze the quark dynamics, it is advantageous to start with the evolution of the total quark number density per unit transverse area and per unit rapidity given by Eq.~\eqref{def_num}. Figure~\ref{fig:numq} shows this quark number density as a function of time. (The number of stochastic fermion configurations used in this computation is $N_\text{conf}=20$.) One observes two characteristic regimes: As shown in more detail in the inset, initially the quark number density increases very rapidly showing a strongly nonlinear behavior which even turns over with a slight intermediate reduction. At later times $Q_s \tau \agt 130$, the quark number density increases almost linearly in time. This rate turns out to be consistent with the kinetic theory estimates based on $2\leftrightarrow 2$ scatterings as indicated by the dashed line in Fig.~\ref{fig:numq}. In the following, we first analyze the earlier-time behavior and then discuss the later times together with the comparison to kinetic theory.

\begin{figure}[t]
 \begin{tabular}{cc}
 \begin{minipage}{0.5\hsize}
 \begin{center}
  \includegraphics[clip,width=7.8cm]{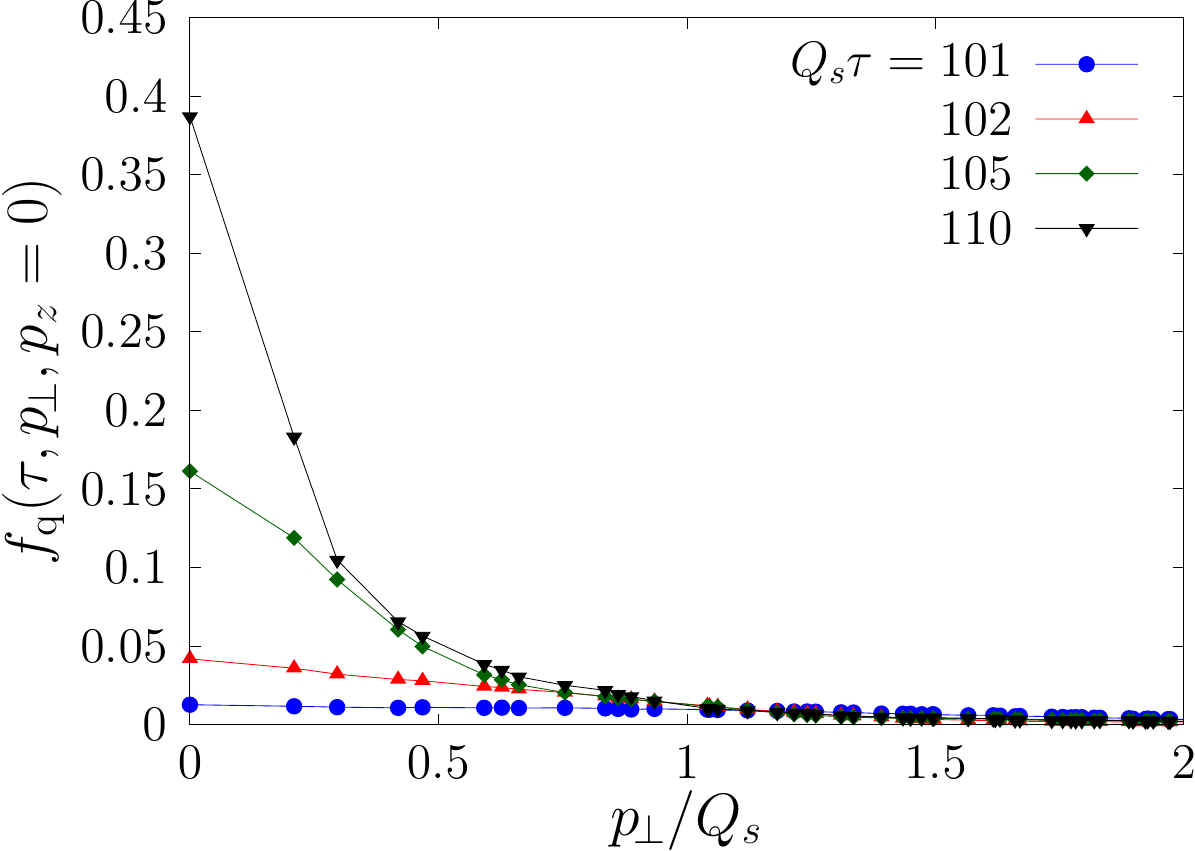} 
 \end{center}
 \end{minipage} &
 \begin{minipage}{0.5\hsize}
 \begin{center}
  \includegraphics[clip,width=7.8cm]{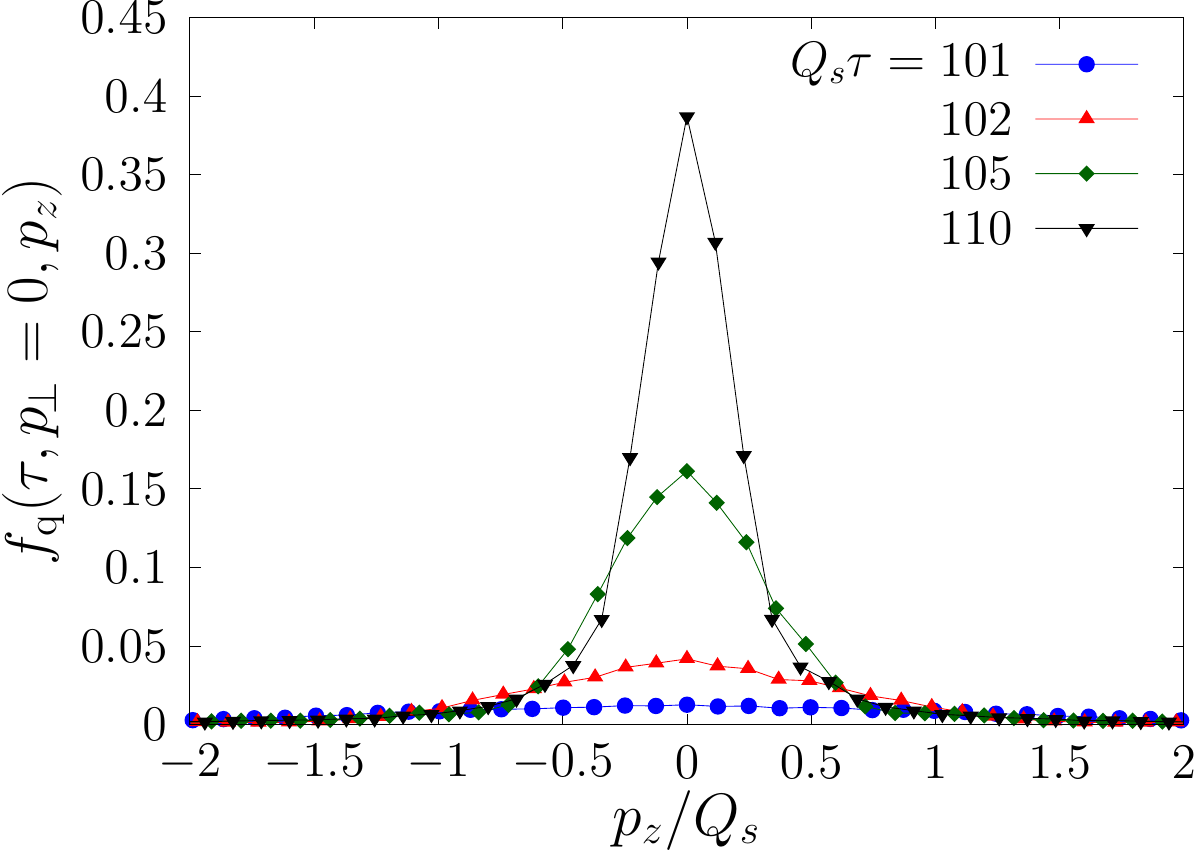} 
 \end{center}
 \end{minipage}
 \end{tabular}
  \caption{Time evolution of the quark distribution function at earlier times $Q_s \tau \leq 110$.
Left: The transverse momentum distribution at $p_z=0$. Right: The longitudinal momentum distribution at $\pperp=0$.}
  \label{fig:distri_early}
\end{figure}

Figure~\ref{fig:distri_early} shows the evolution of the quark momentum distribution function $f_\tq$ during the earlier stage with $Q_s \tau \leq 110$. The left panel displays the transverse momentum distribution evaluated at $p_z=0$, while the right gives the longitudinal momentum distribution at $\pperp =0$. 

\begin{figure}[t]
 \begin{tabular}{cc}
 \begin{minipage}{0.5\hsize}
 \begin{center}
  \includegraphics[clip,width=7.8cm]{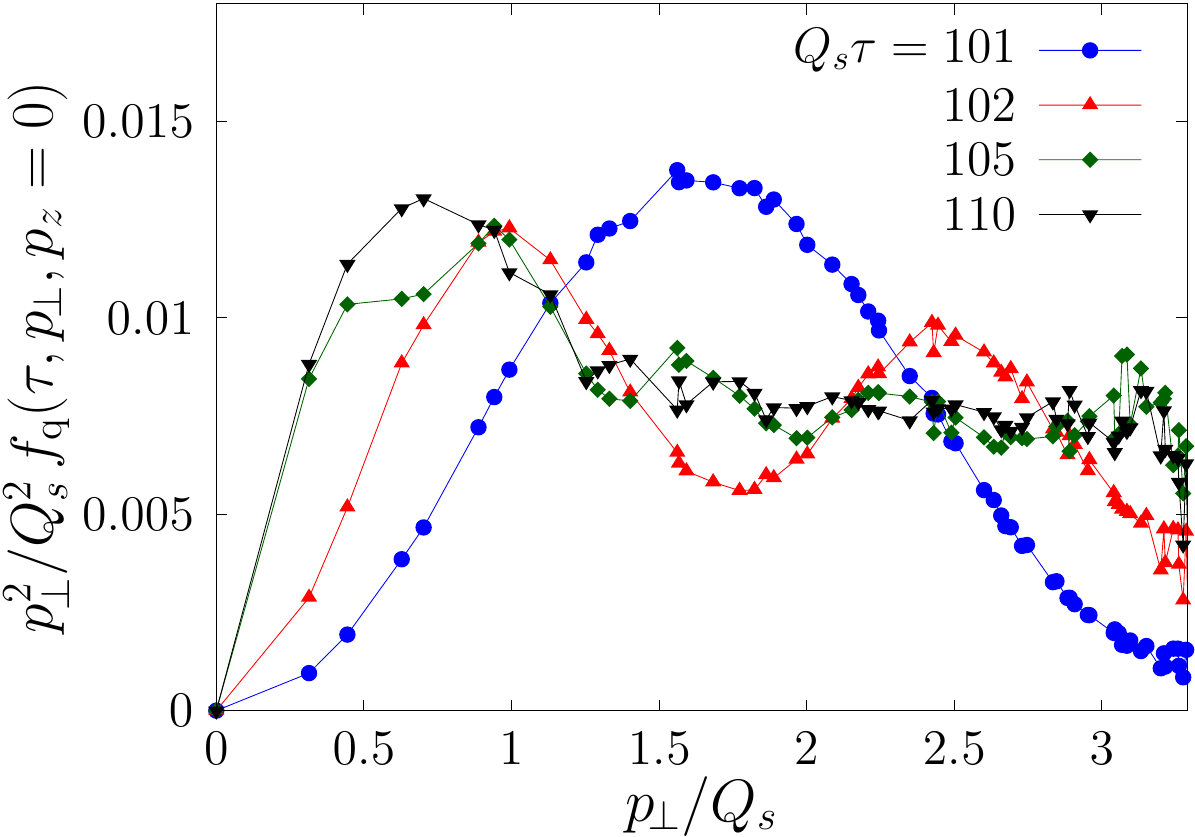} 
 \end{center}
 \end{minipage} &
 \begin{minipage}{0.5\hsize}
 \begin{center}
  \includegraphics[clip,width=7.8cm]{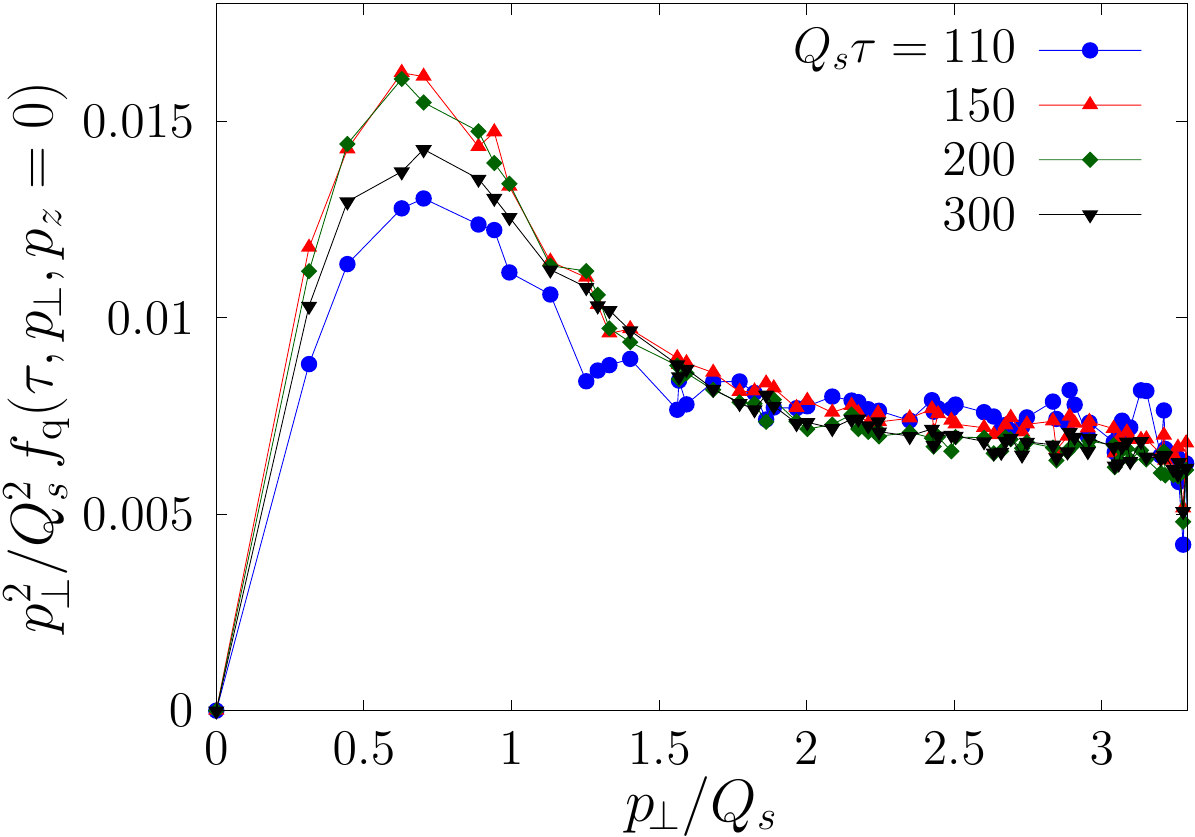} 
 \end{center}
 \end{minipage}
 \end{tabular}
 \caption{The second moment of the quark transverse momentum distribution at earlier times $Q_s \tau \leq 110$ (left) and at later times $Q_s \tau \geq 110$ (right). (Here we used a smaller transverse lattice spacing $Q_s a_\perp =0.417$ in order to have better resolution in the high momentum region.)}
  \label{fig:distri_m}
\end{figure}

Since this early evolution stage far from equilibrium may be expected to be not amenable to a simple quasi-particle picture, we will analyze this in more detail. At very early times $Q_s (\tau -\tau_0) \alt 1$, the occupation number is still much smaller than one. However, the integrated total particle number shows a rapid increase in this time range as shown in Fig.~\ref{fig:numq}. These seemingly inconsistent observations can be understood as a consequence of the fact that the quark spectrum at very early times is dominated by higher momentum modes. To confirm this, we show in Fig.~\ref{fig:distri_m} the second moment of the transverse distribution $\pperp^2 f_\tq$. Although the distribution itself is much smaller at $Q_s \tau=101$ compared to later times, the second moment at $Q_s \tau=101$ is as large as those at later times, since it is dominated by large momenta. At $Q_s \tau=101$, the peak position of the moment is larger than $Q_s$. As time goes on, the peak position is shifted to lower momenta and eventually stabilizes around $p\sim Q_s$, which is the natural scale of this system. In terms of a quasi-particle picture, this behavior is hard to understand. In the very early stage, it is more adequate to interpret the distribution function defined by Eq.~\eqref{def_fq} as a two-point correlation function of the field $\psi (x)$ rather than the momentum distribution of stable quasi-particles. Since we start with the vacuum initial condition for the quark field and the interaction with the overoccupied gauge field is suddenly turned on at the initial time, nontrivial correlations of the quark field are developed at relatively short length scales first and they extend to larger length scales as time proceeds. After this early-time transient stage, the two-point correlation stabilizes, which results in the steady peak position of the moment at later times. 
Then, in the time range $101\alt Q_s \tau \alt 110$, the occupation number of quarks at low momenta $p\alt 0.5 Q_s$ exhibits significant increase.

\begin{figure}[tb]
 \begin{tabular}{cc}
 \begin{minipage}{0.5\hsize}
 \begin{center}
  \includegraphics[clip,width=7.8cm]{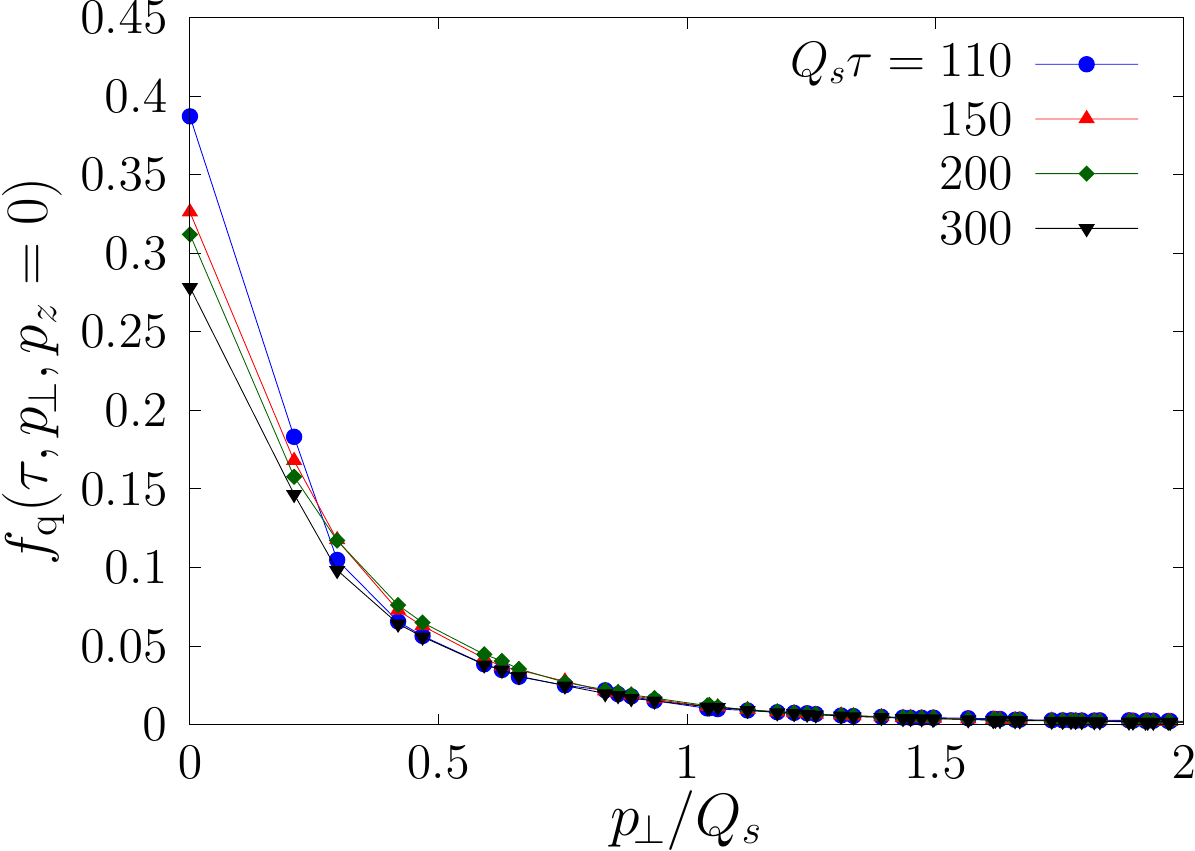} 
 \end{center}
 \end{minipage} &
 \begin{minipage}{0.5\hsize}
 \begin{center}
  \includegraphics[clip,width=7.8cm]{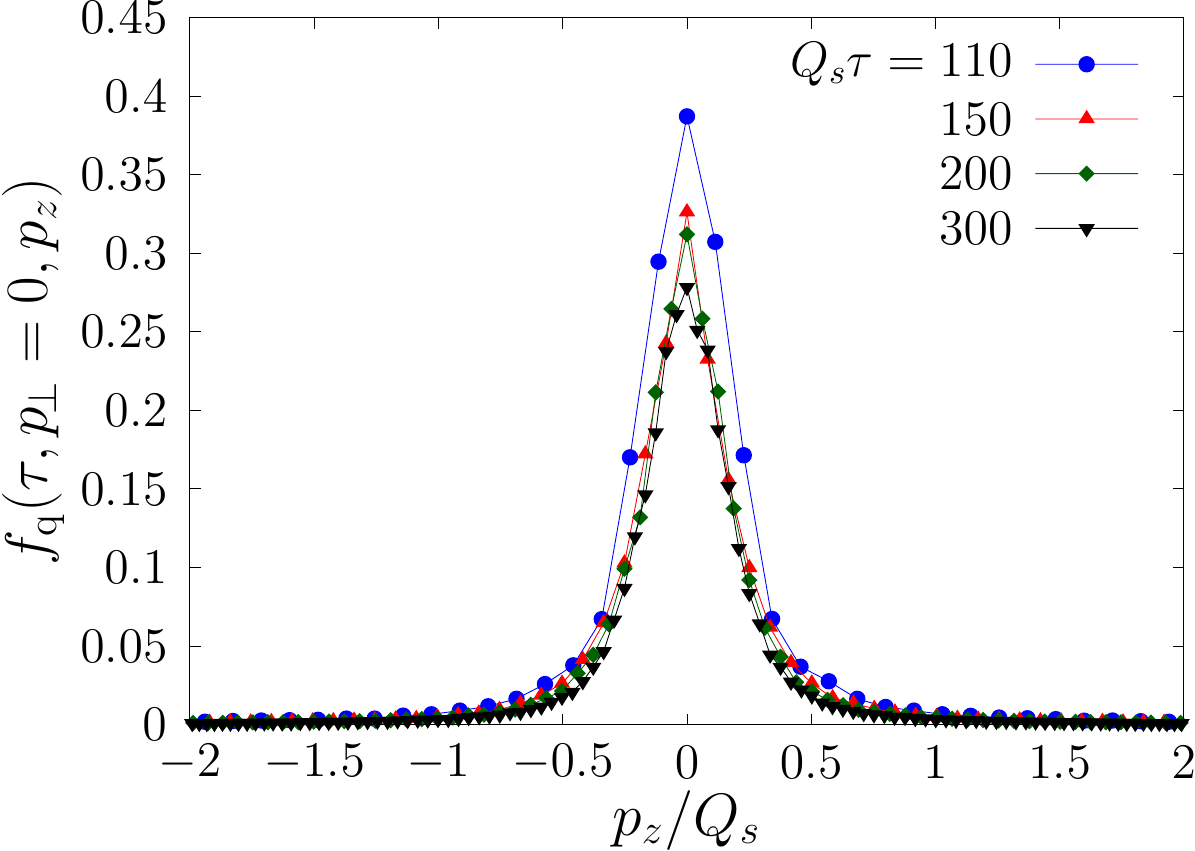} 
 \end{center}
 \end{minipage}
 \end{tabular}
  \caption{Time evolution of the quark distribution function at later times $Q_s \tau \geq 110$.
Left: The transverse momentum distribution at $p_z=0$. Right: The longitudinal momentum distribution at $\pperp=0$.}
  \label{fig:distri_late}
\end{figure}

In Fig.~\ref{fig:distri_late}, we show the time evolution of the quark distribution function at later times $Q_s \tau \geq 110$.
The evolution behavior of the distribution function looks quite different compared to the earlier times shown in Fig.~\ref{fig:distri_early}. The occupation number gradually decreases in time. This behavior can be understood as a consequence of the expansion of the system and the momentum broadening in the longitudinal direction. 
If there is no interaction (free streaming), the transverse distribution would be unchanged, while the width of the longitudinal distribution would shrink in time as $p_z \sim \tau^{-1}$. In Fig.~\ref{fig:distri_late}, the width of the longitudinal distribution  seems to approach a nearly constant value for $Q_s \tau \agt 150$. This means that the longitudinal distribution is broadened by scatterings. For gluons undergoing $2\leftrightarrow 2$ elastic scatterings, the competition between the system's expansion and the scatterings are expected to result in the behavior of the longitudinal momentum as $p_z \sim \tau^{-1/3}$ \cite{Baier:2000sb}. The same behavior can be observed for the quark distribution if the effect of quark production is minor \cite{Tanji:2017suk}. 
The almost constant width of the longitudinal distribution as seen in the right panel of Fig.~\ref{fig:distri_late} indicates the quark production with a nearly constant production rate. Indeed, as shown in Fig.~\ref{fig:numq}, the total quark number density increases almost linearly in this time range.

Remarkably, the increase rate of the total quark number density for $Q_s \tau \agt 130$ turns out to be consistent with the simple kinetic estimate \eqref{kine_nq}. We have inserted the distribution functions obtained from the lattice computations into the kinetic formula \eqref{kine_nq} and integrate it over time. The result is plotted with a dashed line in Fig.~\ref{fig:numq}, where the constant offset, which corresponds to a time-integration constant, has been adjusted to account for the early-time quark production that cannot be described by the simple kinetic theory. The slope of the curve, i.e.\ the production rate, is well reproduced by the kinetic estimate 
with our choice for the cutoff scales in the Coulomb logarithm. 
This result is remarkable because we have taken into account only $2\leftrightarrow 2$ scattering processes in the kinetic estimate. 

We emphasize that one has a priori no reason to expect the kinetic description to 
agree with the lattice results even in the order of magnitude accuracy. 
As shown in Fig.~\ref{fig:distri_gpt}, the occupation number of gluons at their typical momenta is of order $1/g^2$ in this time range. In this case, general $n \leftrightarrow m$ scattering processes that involve arbitrary number of gluons are as important as the $2\leftrightarrow 2$ scattering processes considered in our simple kinetic estimate. In accordance with the earlier findings for the pure gauge theory~\cite{Berges:2013eia,Berges:2013fga}, this observation indicates that the effective kinetic description may be more robust than one can expect based on perturbative power counting. 
We have to note, though, that here we compare only the integrated number density and that we perform a rather indirect comparison by inserting the distribution functions obtained from the lattice simulations into the kinetic formula. 
Furthermore, we have applied the small-angle approximation for the kinetic theory, that involves logarithmic uncertainty. 
To establish a firmer connection between the real-time lattice simulations and the kinetic theory, one has to make a direct comparison for the unintegrated distribution functions 
without employing the small-angle approximation
as was done in Ref.~\cite{York:2014wja} in a non-expanding system. We leave such computations for future investigations. 

\begin{figure}[tb]
 \begin{center}
  \includegraphics[clip,width=8cm]{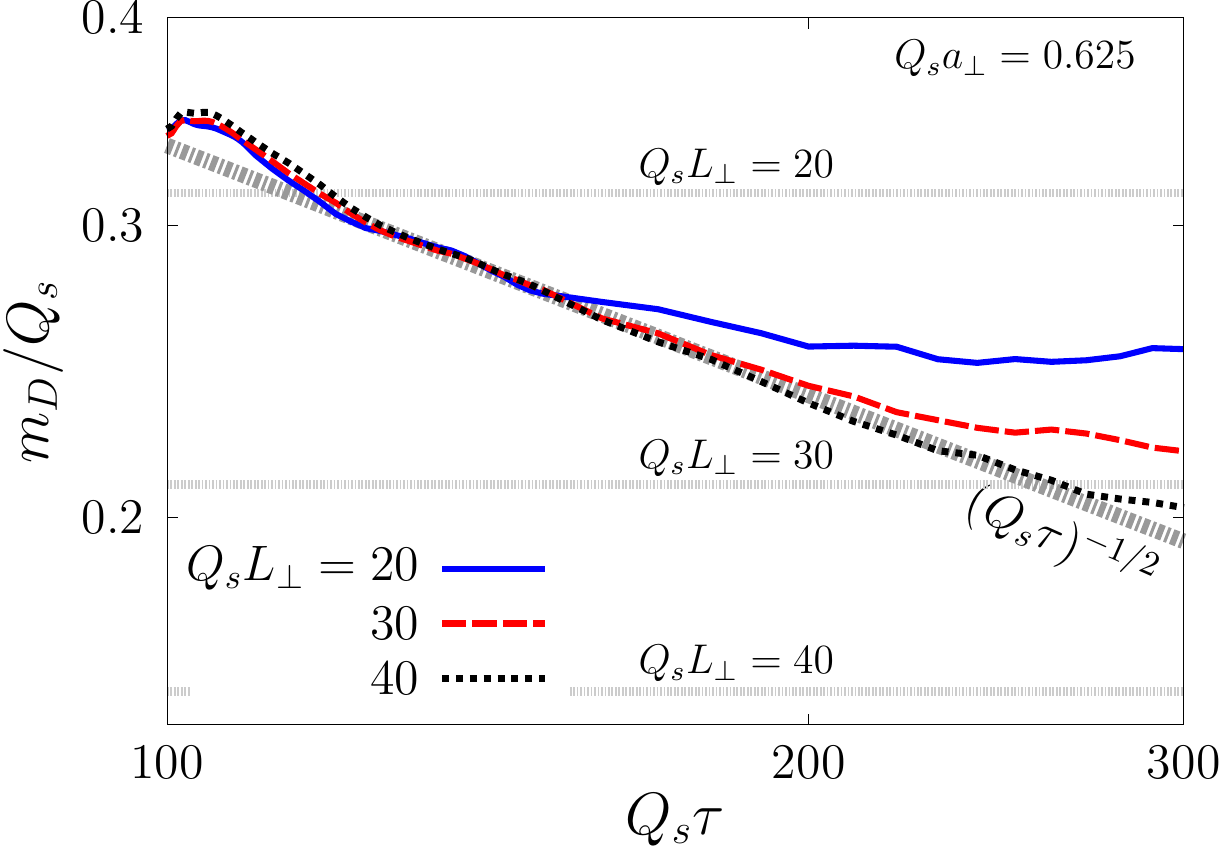}  \vspace{-10pt}
  \caption{Time evolution of the Debye mass scale. Three different transverse system sizes are compared.
  The smallest nonzero transverse momenta are indicated by gray horizontal lines for each system size. The expected temporal power law $\tau^{-1/2}$ is also indicated by a gray dashed line. These results are obtained by pure Yang-Mills simulations with the lattice parameters $Q_s a_\perp =0.625$, $N_\eta=256$, and $a_\eta=1.95 \times 10^{-3}$. }
  \label{fig:Debye1}
 \end{center}
\end{figure}

One may notice that the agreement between the lattice result and the kinetic estimate becomes worse for $Q_s \tau \agt 250$ in Fig.~\ref{fig:numq}. This can be attributed to a lattice artifact rather than the failure of the kinetic description. In the kinetic description discussed in Sec.~\ref{sec:kinetic}, the Debye mass scale $m_D$ plays a crucial role. Indeed, the production rate \eqref{kine_nq} is proportional to $m_D^4$ if we drop the contributions from the quark distribution. Therefore, it is important to precisely resolve the Debye mass scale on the lattice. The time evolution of the Debye mass scale is shown in Fig.~\ref{fig:Debye1} for three different transverse system sizes. Since the contribution from quarks to the Debye mass is negligible for $g^2 N_f \ll 1$, we show results obtained by pure Yang-Mills simulations. As long as the gluon distribution is dominated by the hard scale $Q_s$, we can expect that the Debye mass scale decreases in time as $\tau^{-1/2}$ except at very early times. This behavior has been confirmed by large-scale lattice gauge theory simulations \cite{Berges:2013fga} and kinetic theory calculations \cite{Tanji:2017suk}. For the smallest system size ($Q_s L_\perp =20$), hence for the coarsest resolution in the infrared, significant deviations from the behavior $m_D \sim \tau^{-1/2}$ is seen. For the largest system size ($Q_s L_\perp =40$), the temporal evolution of the Debye mass is better resolved. Because of the significant numerical costs of the quark sector, we have used the transverse system size $Q_s L_\perp =30$ in the computations with quarks. Therefore, the infrared sector is not sufficiently well resolved at later times. This is likely to be the reason for the discrepancy between the lattice result and the kinetic estimate that appears for $Q_s \tau \agt 250$ in Fig.~\ref{fig:numq}. 

\begin{figure}[tb]
 \begin{center}
  \includegraphics[clip,width=8cm]{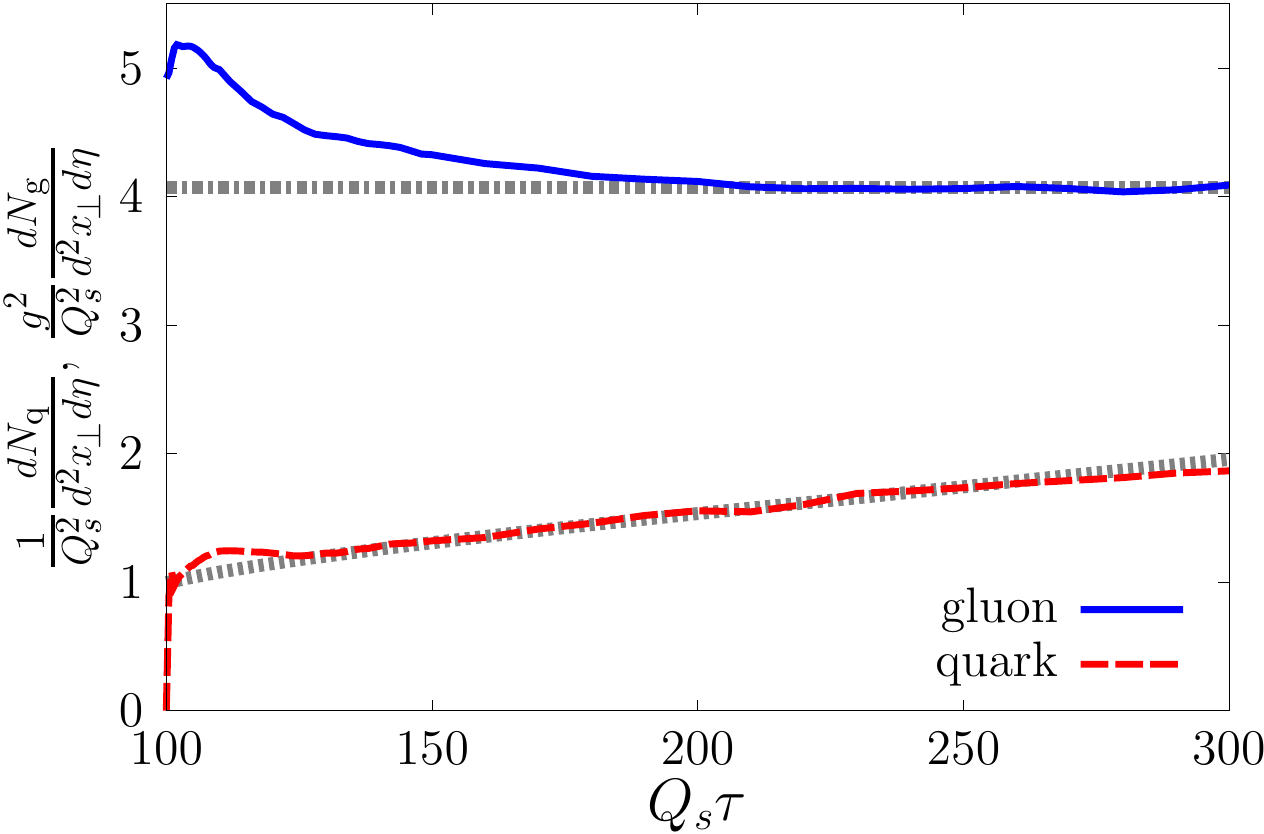}  \vspace{-10pt}
  \caption{Time evolution of the number densities of gluons and quarks. The gluon number is multiplied by $g^2=10^{-4}$. 
The kinetic theory estimates are indicated by gray dashed lines.}
  \label{fig:numcomp}
 \end{center}
\end{figure}

To gain more insight into physical processes realized in the real-time lattice simulations, we make a comparison to the kinetic theory estimate also in the gluon sector. In Fig.~\ref{fig:numcomp}, the total gluon number density per unit transverse area and per unit rapidity is plotted as a function of time, as well as the total quark number density. Since the occupation number of gluons is the order of $1/g^2$, which is much larger than that of quarks, we multiply the factor $g^2=10^{-4}$ to the gluon number density. The kinetic theory estimates are indicated by gray dashed lines. For $1/g^2 \gg 1$, the decrease of the gluon number by the kinetic $2\leftrightarrow 2$ scattering process is negligible. Therefore, the kinetic estimate for the gluon number gives essentially a horizontal line in this figure. Indeed, the gluon number obtained from the lattice simulations seems to approach a constant value. This observation indicates that the dynamics of the overoccupied gluon plasma at weak coupling is dominated by the elastic $2\leftrightarrow 2$ scatterings among gluons, and reinforces the observations made in previous studies \cite{Berges:2013eia,Berges:2013fga,Tanji:2017suk}. 
Compared to quarks, the time at which the kinetic estimate starts to agree well with the lattice result appears to be somewhat later for gluon. Being fermions, quarks respect the exclusion principle which might help in behaving more like quasi-particles. 

\begin{figure}[tb]
 \begin{center}
  \includegraphics[clip,width=8cm]{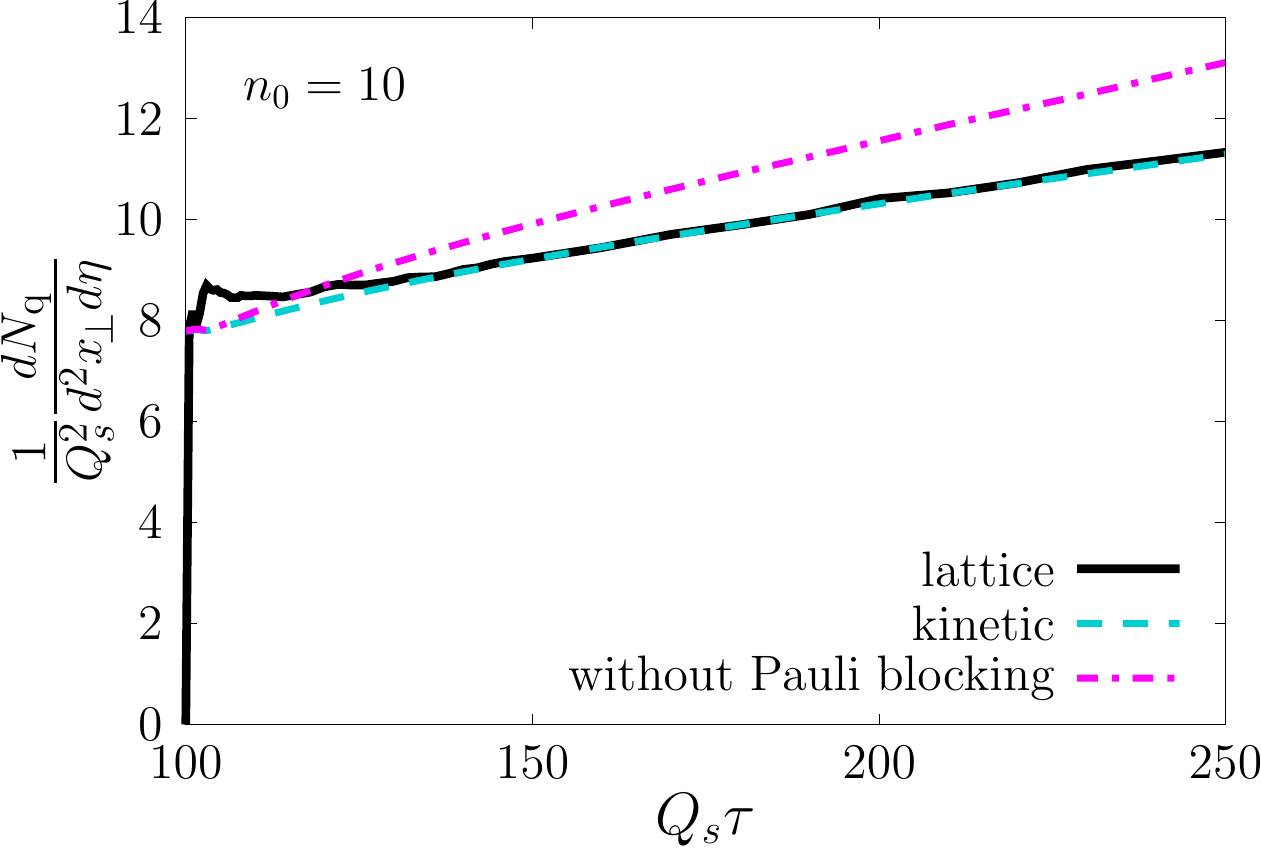} \vspace{-10pt}
  \caption{Quark number density for $n_0=10$. 
The kinetic estimate \eqref{kine_nq} well fits the lattice result. Without the Pauli blocking term, the production rate is overestimated. }
  \label{fig:n0_numq}
 \end{center}
\end{figure}

So far, we have fixed the initial occupation parameter for gluons to be $n_0=1$. 
Before closing this subsection, we discuss a result with larger $n_0$. Figure~\ref{fig:n0_numq} shows the time evolution of the total quark number density for $n_0=10$. Compared to the case of smaller gluon occupancies, the quark production is more efficient as expected. However, the produced quark number is not linear in $n_0$ because of Pauli's exclusion principle. 
Also for $n_0=10$, the kinetic theory estimate \eqref{kine_nq} well reproduces the quark production rate. This is somewhat surprising because for too large gluon occupancies the kinetic theory description is expected to become unreliable. Our results indicate that the simple $2\leftrightarrow 2$ kinetic theory estimate for the quark production rate is quite robust.
In the figure, we also plot the kinetic estimate after dropping the Pauli blocking factor $(1-2f_\tq)$ in Eq.~\eqref{kine_nq}. Without this term, the kinetic theory overestimates the lattice result for quark production. Since the quark occupation number of order one is developed in the time range $101 \alt Q_s \tau \alt 110$, the kinetic quark production at later times is affected by Pauli blocking.

\subsection{Quark mass dependence} \label{subsec:mass}

\begin{figure}[tb]
 \begin{center}
  \includegraphics[clip,width=8cm]{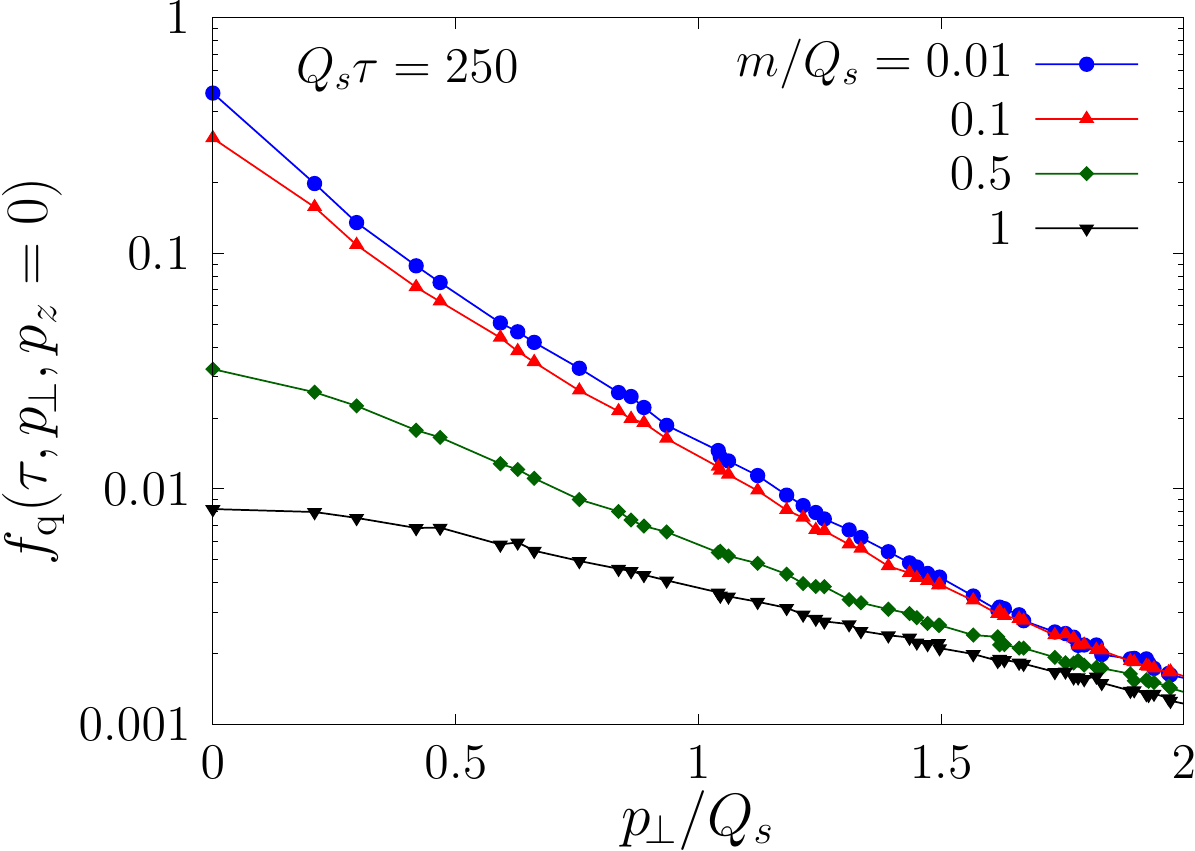} \vspace{-10pt}
  \caption{Quark transverse momentum distribution for different quark masses at the time $Q_s \tau=250$. }
  \label{fig:massdep_pt}
 \end{center}
\end{figure}

\begin{figure}[tb]
 \begin{center}
  \includegraphics[clip,width=8cm]{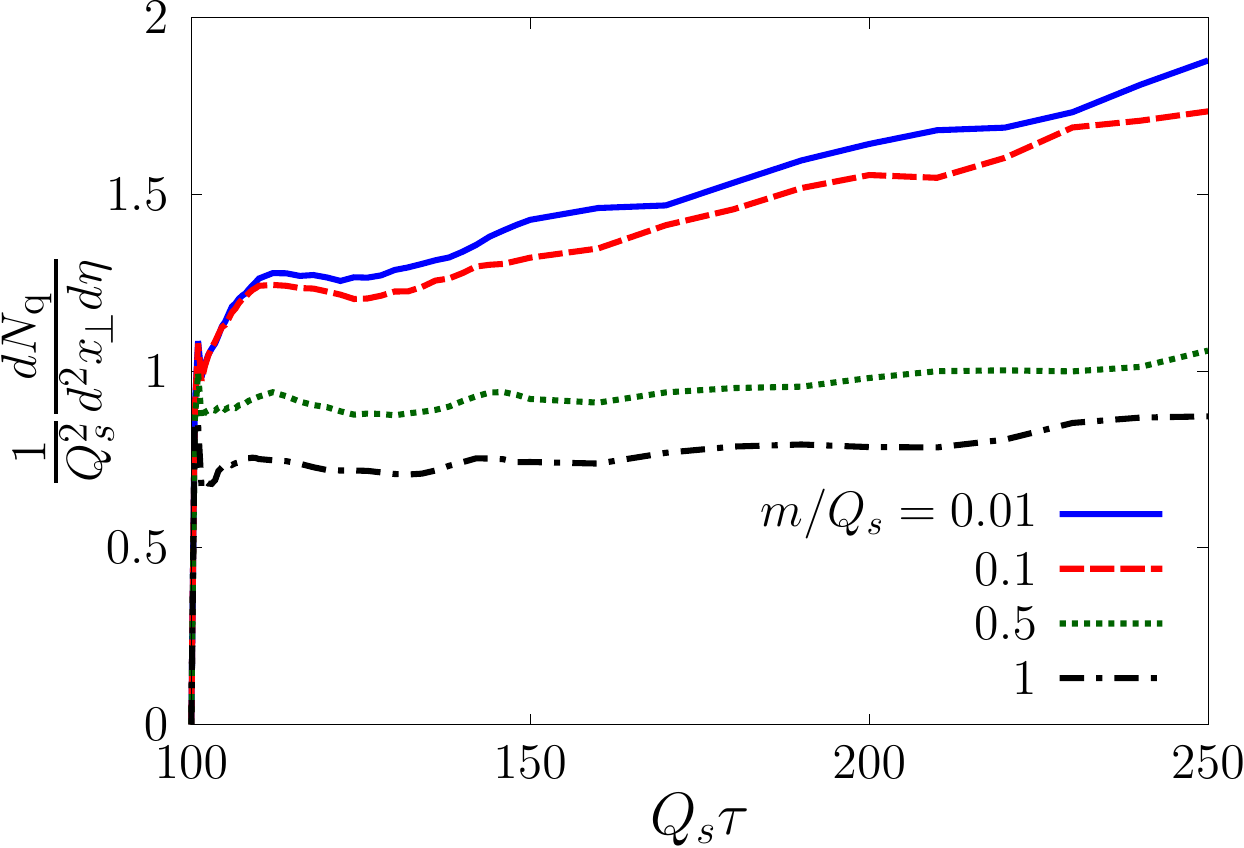} \vspace{-10pt}
  \caption{Time evolution of the quark number density for different quark masses.}
  \label{fig:massdep_numq}
 \end{center}
\end{figure}

In this subsection, we investigate the quark mass dependence of our lattice simulation results.
We still keep the number of fermion flavor to be $N_f=1$. We note that as long as $g^2 N_f \ll 1$ the effect of backreactions from quarks onto gluons is minor and, therefore, the corresponding results are linear in $N_f$. 

We compare different quark masses in units of the saturation scale, $m/Q_s=0.01$, $0.1$, $0.5$, and~$1$. 
For $Q_s \simeq 1$ GeV, up and down quark masses are of the order of or lighter than $m/Q_s=0.01$, the strange quark mass is of the order of $m/Q_s=0.1$, and the charm quark mass is of the order of $m/Q_s=1$. 
In Fig.~\ref{fig:massdep_pt}, the quark transverse momentum distribution evaluated at $p_z=0$ and $Q_s \tau=250$ is plotted for the different quark masses. 
The production of heavy quarks is suppressed and the natural mass ordering becomes apparent. 
The total quark number density is plotted as a function of time in Fig.~\ref{fig:massdep_numq} for the different masses. The quark production rate especially at later times is lower for heavier quarks. We have confirmed that the kinetic theory estimate agrees well with the lattice results for $m/Q_s =0.01$ and $m/Q_s =0.1$. The small-angle approximation is not expected to apply to larger quark masses, since it is valid only if the mass of an exchanged particle, i.e.~quark in the present case, is negligible compared to the typical scale of the distributions. Indeed we even find that the small-angle approximation overestimates the lattice results for heavier masses $m/Q_s =0.5$ and $1$, which we attribute to the fact that the approximation is used beyond its range of validity. We leave the important task of a comparison to kinetic theory without employing the small-angle approximation for future investigations.

\begin{figure}[tb]
 \begin{tabular}{cc}
 \begin{minipage}{0.5\hsize}
  \begin{center}
   \includegraphics[clip,width=7.8cm]{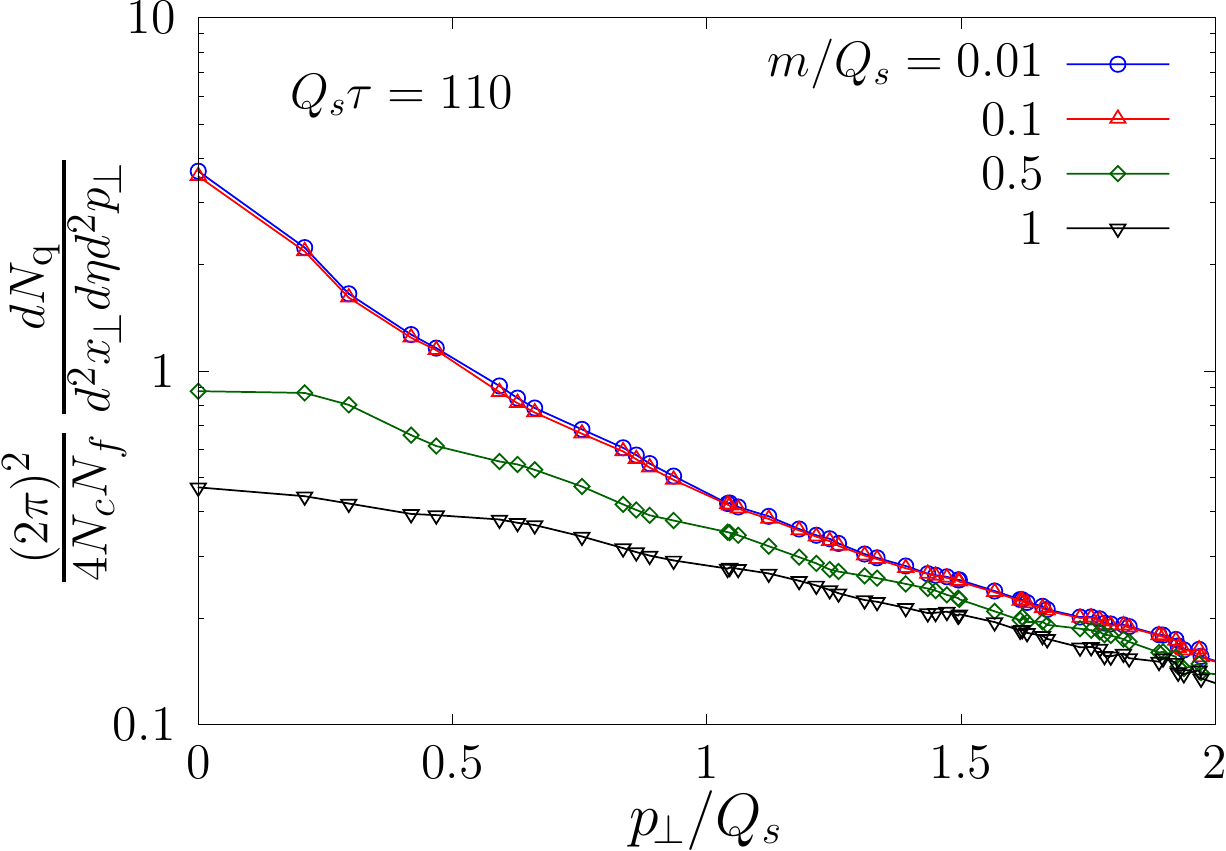}
  \end{center}
 \end{minipage} &
 \begin{minipage}{0.5\hsize}
  \begin{center}
   \includegraphics[clip,width=7.8cm]{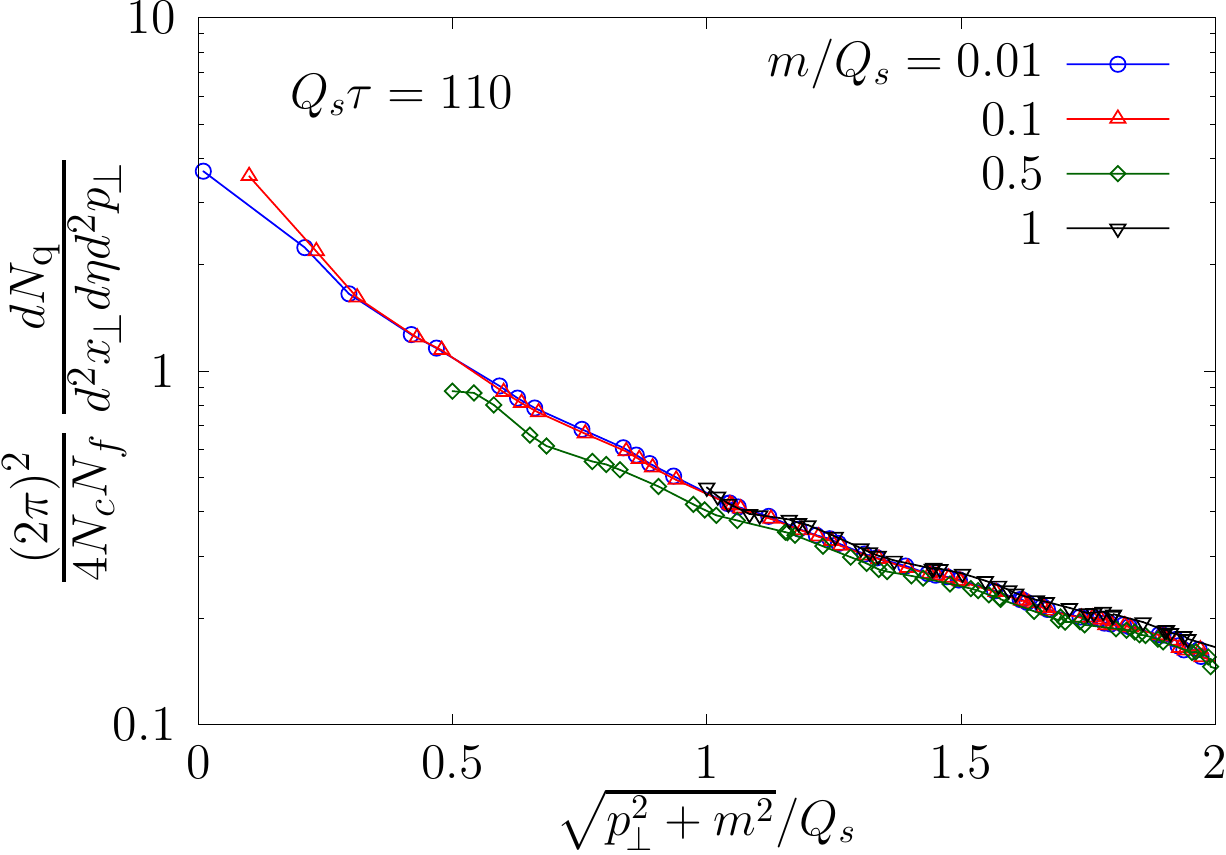}
  \end{center}
 \end{minipage} 
 \end{tabular}
\caption{The integrated transverse spectrum for different quark masses at $Q_s \tau=110$ as a function of the transverse momentum (left) and of the transverse mass (right). 
}
\label{fig:massdep_pt_int1}
\end{figure}
\begin{figure}[tb]
 \begin{tabular}{cc}
 \begin{minipage}{0.5\hsize}
  \begin{center}
   \includegraphics[clip,width=7.8cm]{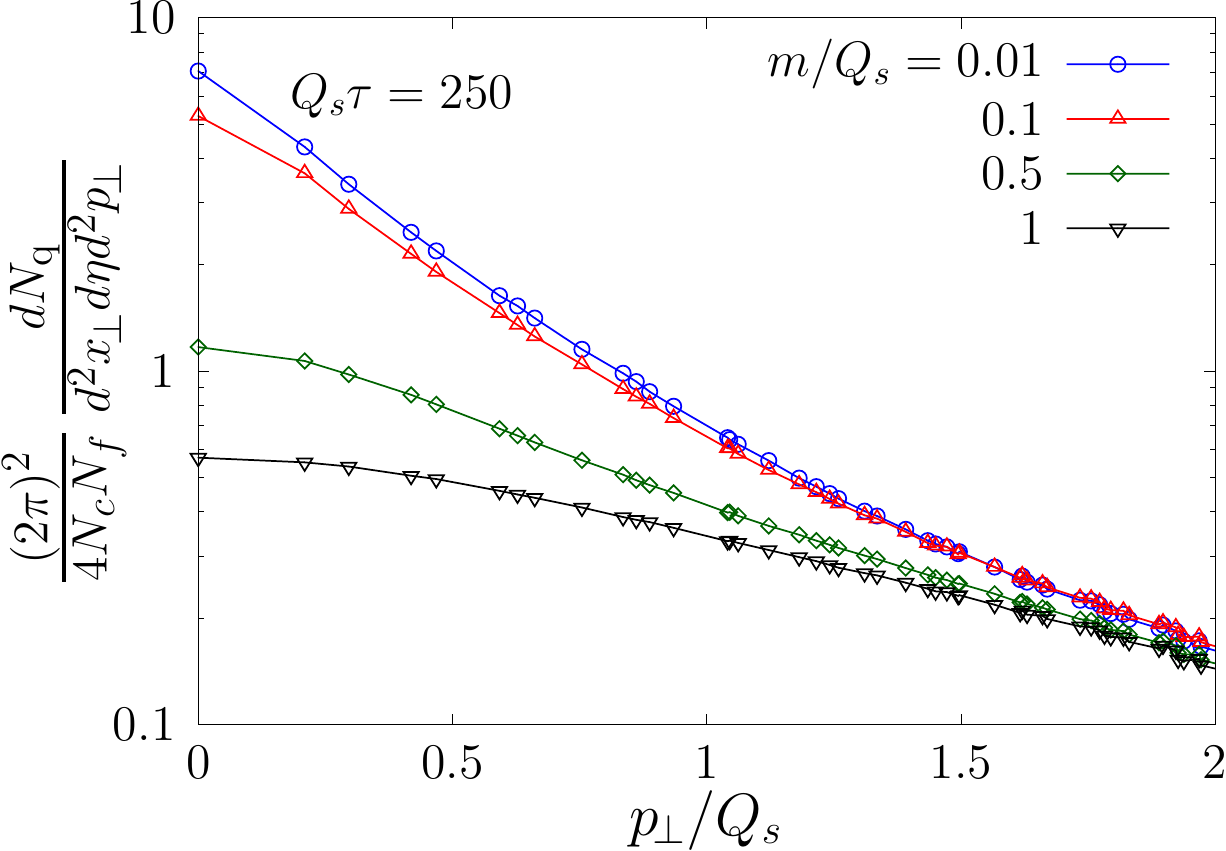}
  \end{center}
 \end{minipage} &
 \begin{minipage}{0.5\hsize}
  \begin{center}
   \includegraphics[clip,width=7.8cm]{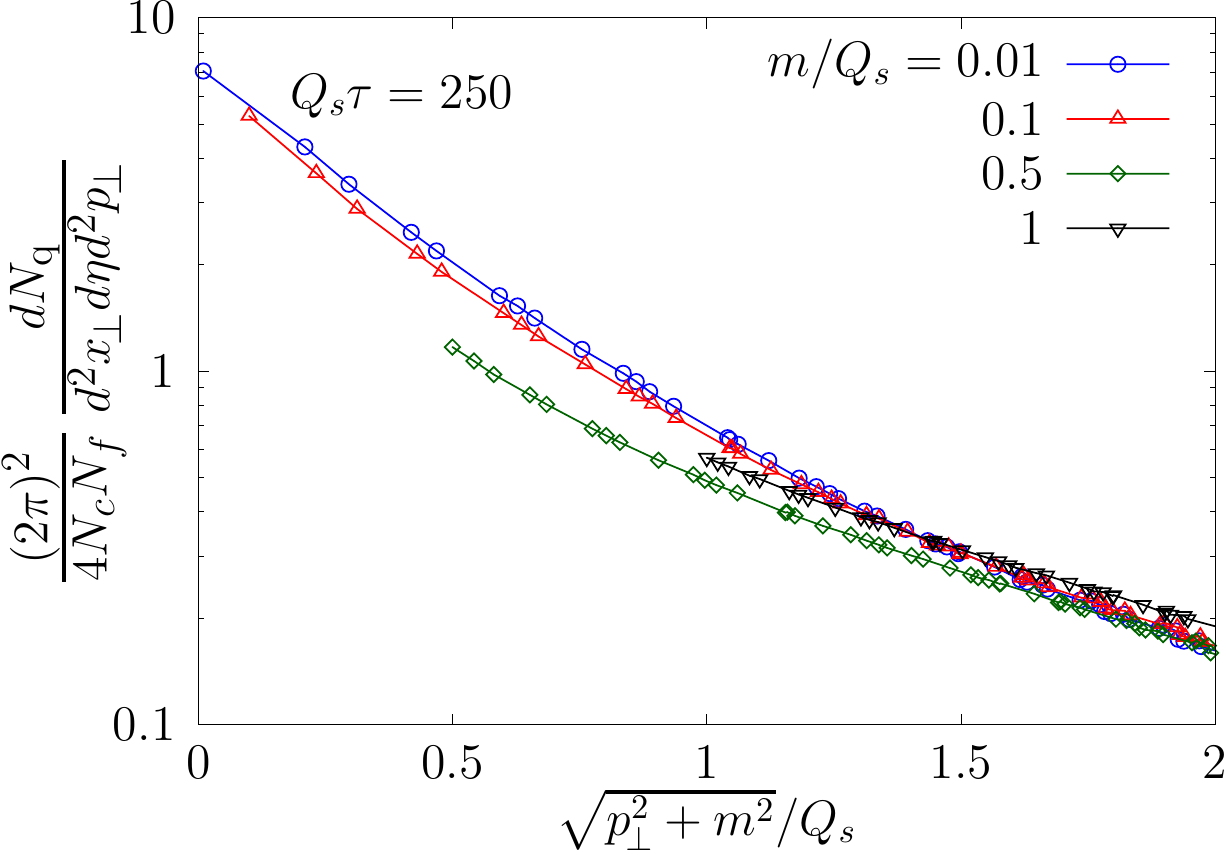}
  \end{center}
 \end{minipage} 
 \end{tabular}
\caption{The integrated transverse spectrum for different quark masses at $Q_s \tau=250$ as a function of the transverse momentum (left) and of the transverse mass (right).}
\label{fig:massdep_pt_int2}
\end{figure}

In the $\tau$-$\eta$ coordinates, $p_z=\nu /\tau$ denotes the longitudinal momentum measured in the co-moving frame with velocity $v_z =\text{tanh} (\eta)$ \cite{Tanji:2010ui}. Therefore, it is not directly measurable in experiments. 
The transverse momentum spectrum integrated over longitudinal momentum would be of more direct experimental interest. It can be computed from the distribution function $f_\tq (\tau ,\bpperp ,p_z )$ as
\begin{equation}
(2\pi)^2 \frac{dN_\tq}{d^2 x_{\!\perp} d\eta d^2 p_{\!\perp}}
= \frac{4N_c N_f}{L_\eta} \sum_\nu \, f_\tq (\tau ,\bpperp ,\nu/\tau) \, .
\end{equation}
In the left panels of Figs.~\ref{fig:massdep_pt_int1} and \ref{fig:massdep_pt_int2}, this quantity is plotted as a function of the transverse momentum for different quark masses at times $Q_s \tau=110$ and $250$, respectively.  
The same mass ordering as seen in the transverse momentum distribution at $p_z=0$ is observed. 
In the right panels, the same quantity is plotted as a function of transverse mass $m_\perp =\sqrt{\pperp^2 +m^2}$. 
Interestingly, at $Q_s \tau=110$, all points for different masses lie on top of each other in the region $\sqrt{\pperp^2 +m^2} \agt Q_s$. This means that the quark transverse spectrum in this region is only a function of transverse mass.

At a later time, $Q_s \tau=250$, the transverse-mass scaling becomes less accurate as shown in the right panel of Fig.~\ref{fig:massdep_pt_int2}. This observation indicates that the scaling is valid only for the early-time nonperturbative quark production, which appears not to be described by the kinetic theory.\footnote{We note that nonperturbative particle production by the Schwinger mechanism satisfies the same scaling.} In contrast, the perturbative scattering amplitude, which is encoded in the collision term of the kinetic equation, generally does not satisfy the transverse-mass scaling as it depends on $\bpperp$ and $m$ individually. Since the later-time quark production is well described by the $2\leftrightarrow 2$ kinetic process, the transverse-mass scaling is broken at later times.

\subsection{Large $N_f$} \label{subsec:largeNf}

So far, we have studied $N_f=1$ at weak coupling $g=10^{-2}$. In this case, the backreaction from quarks to gauge fields has little effects. While the real-time lattice simulation method requires a weak coupling, by considering large $N_f$, such that $g^2 N_f$ becomes sizable, we can investigate the impact of the backreaction of the produced quarks onto the gluons~\cite{Gelfand:2016prm}.

In the following, we show results for $g^2 N_f =0.5$. 
Since the statistical fluctuations in the color current \eqref{current} are amplified by the factor $N_f$, the number of the stochastic fermion fields $N_\text{conf}$ needs to be larger than for the case of small $N_f$ in order to achieve the same accuracy. We use $N_\text{conf}=480$, with which we have confirmed a good convergence of the results. Because  the numerical costs are high for large $N_\text{conf}$, we have employed a smaller lattice size, $N_\perp =32$, $N_\eta =256$, $Q_s a_\perp =0.625$, and $a_\eta =1.95 \times 10^{-3}$. 

\begin{figure}[tb]
 \begin{tabular}{cc}
 \begin{minipage}{0.5\hsize}
  \begin{center}
   \includegraphics[clip,width=7.8cm]{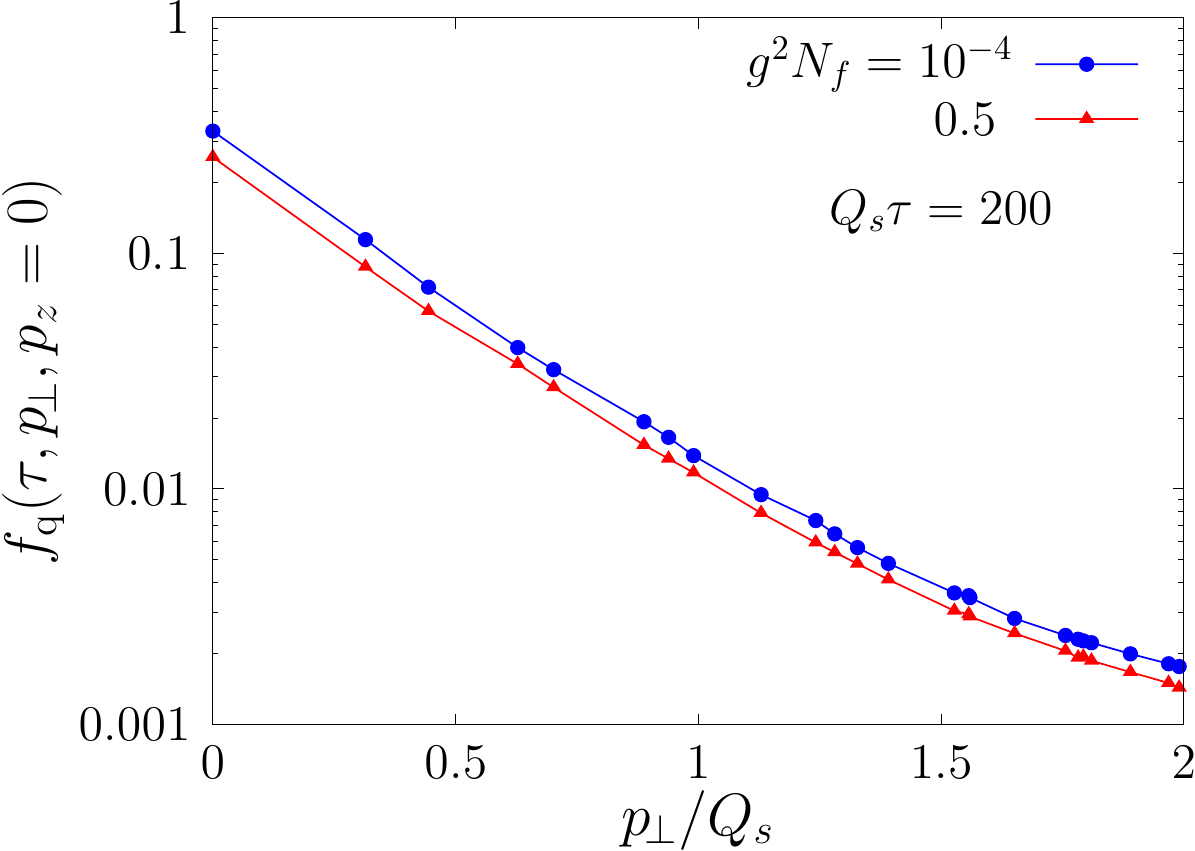}
  \end{center}
 \end{minipage} &
 \begin{minipage}{0.5\hsize}
  \begin{center}
   \includegraphics[clip,width=7.8cm]{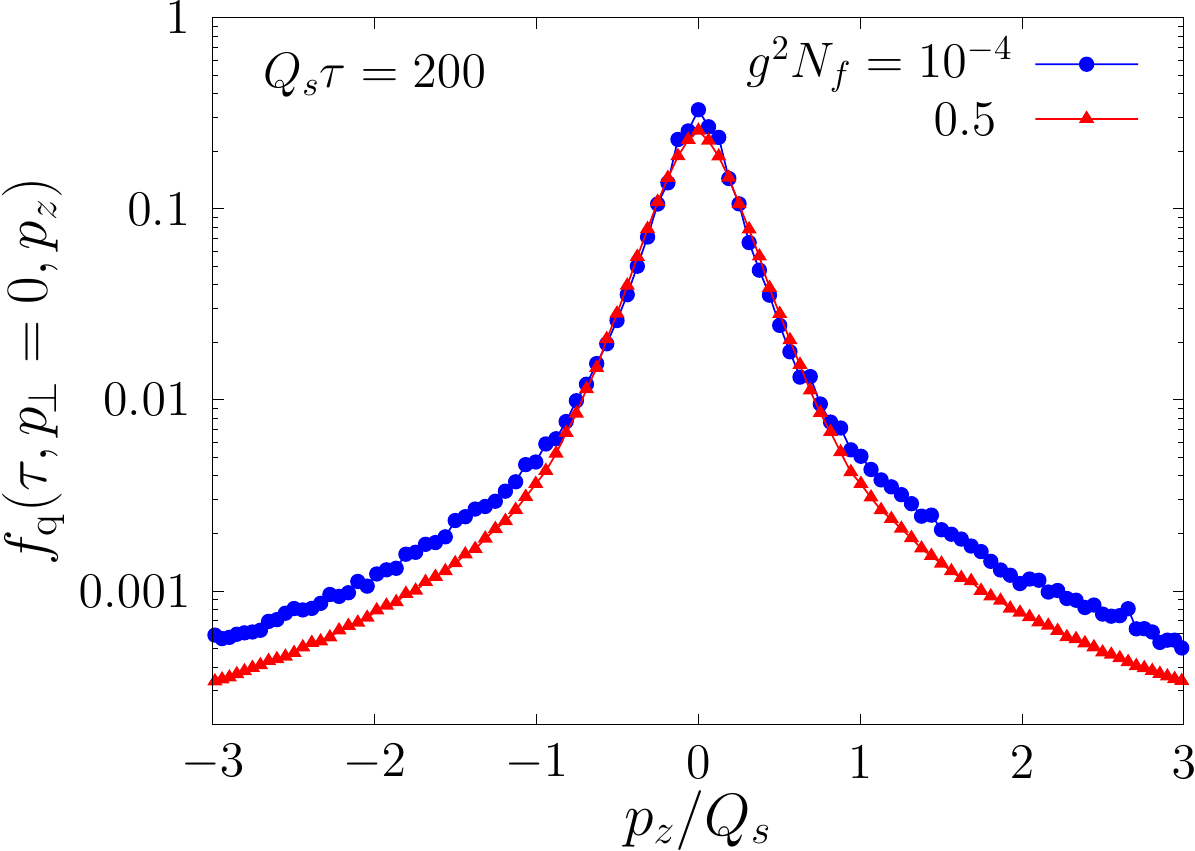}
  \end{center}
 \end{minipage} 
 \end{tabular}
\caption{Quark distribution functions at $Q_s \tau =200$ for $g^2N_f=0.5$ as compared to $g^2 N_f=10^{-4}$.
Left: Transverse momentum dependence at $p_z=0$. Right: Longitudinal momentum dependence at $\pperp =0$.}
\label{fig:Nfdep_distri}
\end{figure}

\begin{figure}[tb]
 \begin{center}
  \includegraphics[clip,width=8cm]{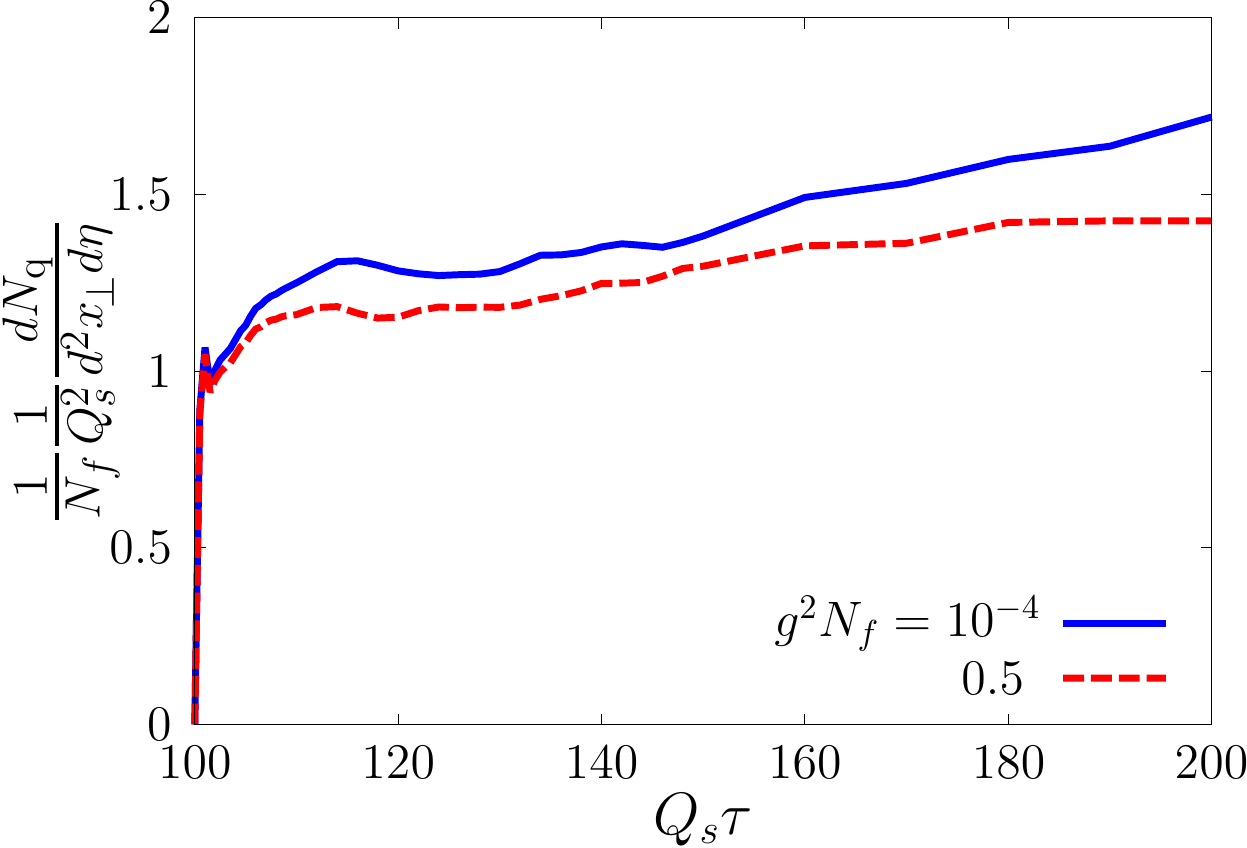} 
  \caption{Quark number density per flavor for $g^2N_f=0.5$ as compared to $g^2 N_f=10^{-4}$.}
  \label{fig:Nfdep_numq}
 \end{center}
\end{figure}

In Fig.~\ref{fig:Nfdep_distri}, we compare the quark one-particle distribution function for $g^2 N_f=0.5$ to that for $g^2 N_f =10^{-4}$. 
With larger $N_f$, the distribution decreases slightly. This is a result of the fact that the gauge fields are more diminished by the backreaction from the quarks for larger $N_f$, and hence the quark production per flavor gets less efficient.
We can observe the same tendency in Fig.~\ref{fig:Nfdep_numq}, where the quark number density per  flavor is compared for $g^2 N_f=0.5$ and $10^{-4}$.  
At very early times, the difference between the two cases is not significant because the effects of the backreaction is still minor then. As time goes on, the effects of the backreaction  accumulate and the difference becomes visible. 

\begin{figure}[tb]
 \begin{center}
  \includegraphics[clip,width=8cm]{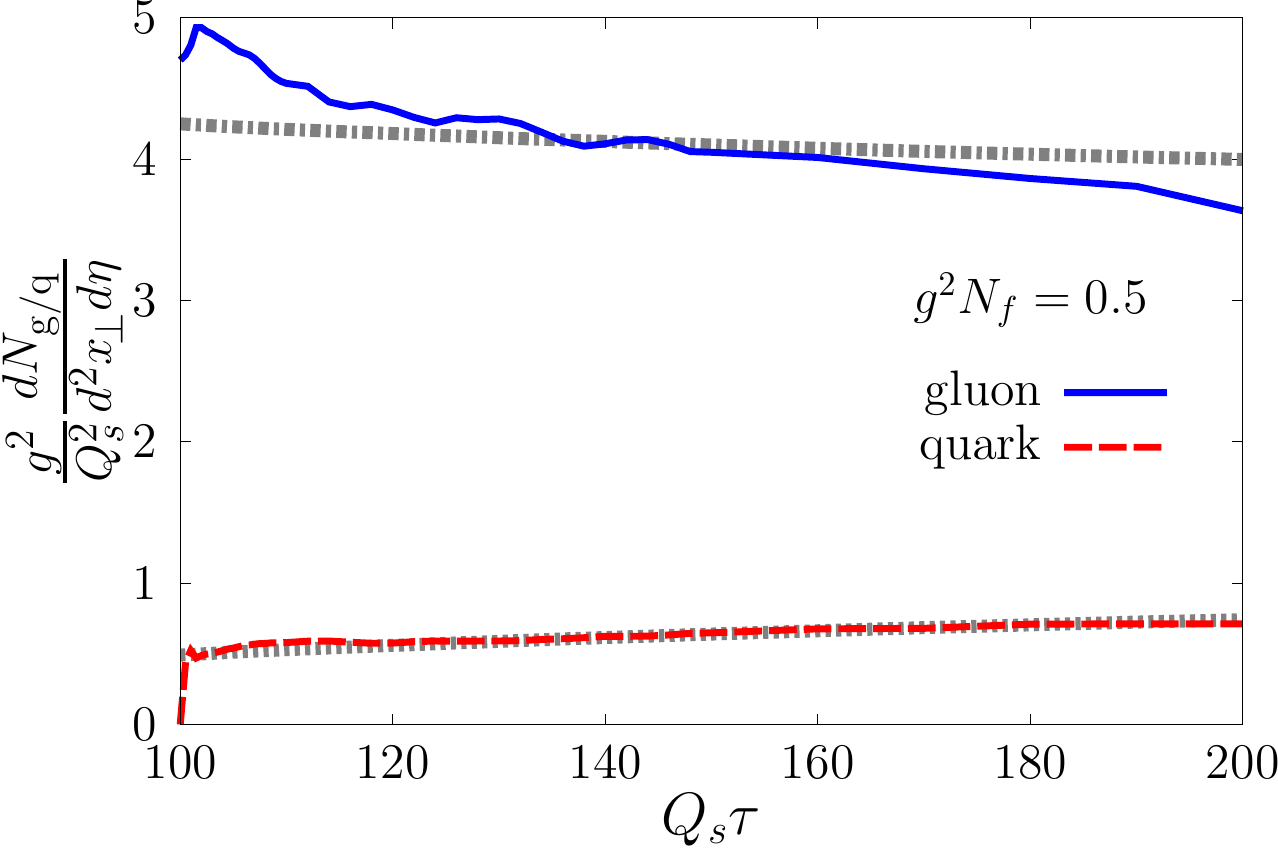} 
  \caption{Comparison of the gluon number density and the quark number density for $g^2N_f=0.5$.
  Both of the number densities are multiplied by $g^2=10^{-4}$. 
  The kinetic theory estimates based on Eq.~\eqref{kine_nq} are indicated by gray dashed lines.}
  \label{fig:Nfdep_numcomp}
 \end{center}
\end{figure}

We find that the later-time quark production is still consistent with the kinetic theory description also for the case of large $N_f$. 
In Fig.~\ref{fig:Nfdep_numcomp}, the number densities of quarks and gluons for $g^2 N_f=0.5$ are plotted as a function of time. Both of the number densities are multiplied by the factor $g^2=10^{-4}$. Similarly to Fig.~\ref{fig:numcomp}, the kinetic theory estimates based on Eq.~\eqref{kine_nq} are indicated by gray dashed lines. 
For large $N_f$, the Debye mass scale and the kinetic production rate \eqref{kine_nq} have large contribution from the quark sector. It seems highly non-trivial that the production rate for quarks still shows a remarkably good agreement with the simple kinetic estimate for this strongly correlated case. In contrast, the decrease of the gluon number exhibits sizable deviations from the kinetic estimate.

\section{Conclusions} \label{sec:conclusion}

We have studied the nonequilibrium time evolution of the longitudinally expanding non-Abelian plasma with two colors and $N_f$ degenerate quark flavors of mass $m$. While the initially highly occupied gluons at weak coupling are found to lead to strongly enhanced quark production, the backreaction of the quark sector on the gluon distribution has only minor consequences for small $g^2 N_f$ at not too late times. In contrast, the backreaction of quarks becomes sizable for $g^2 N_f$ approaching order one. In view of applications to heavy-ion collisions, taking $g^2 N_f$
of order one is expected to be a reasonable assumption and our results provide valuable insights into real-time QCD dynamics from first principles.  

We find that the nonequilibrium quark production is characterized -- after a short period of an initial rapid increase -- by an almost linear growth of total quark number with time. The same qualitative behavior is observed for a wide range of $g^2 N_f$ and quark masses employed, while the production rate of quark number density per flavor decreases for larger $g^2 N_f$ because of a reduction of scattering rates by the diminished gluon occupation numbers for enhanced quark backreactions.

Remarkably, for not too large quark masses the linear growth of the total quark number appears to be consistent with a simple kinetic estimate including only $2 \leftrightarrow 2$ scatterings in the small-angle approximation. We emphasize that there is no reason to expect this a priori in a far-from-equilibrium regime of highly occupied gluons and, in particular, even for a strongly correlated quark sector where $g^2 N_f$ is not small. This apparent `unreasonable effectiveness' of effective kinetic descriptions, at least for some important aspects of the nonequilibrium dynamics, 
still poses major open questions that require further studies featuring a direct solution of kinetic theory with quarks. While this is beyond the scope of the real-time lattice simulation study we have focused on in this work, hopefully the first-principles lattice results help to shed some light on the problem of finding a consistent approximation scheme justifying employed effective kinetic descriptions for the thermalization dynamics of heavy-ion collisions. 

By investigating the quark production for different quark masses, we have also found that quarks produced in the early-time nonlinear regime satisfy a nonequilibrium scaling law, namely `transverse-mass' scaling. The transverse momentum spectra for a wide range of quark mass depend on transverse momentum and quark mass only through the transverse mass term $\sqrt{\pperp^2 +m^2}$. Since this scaling law is characteristic for the early-time nonperturbative quark production, it may help differentiating the quark production mechanisms in the dynamical evolution history of a heavy-ion collision. 

\section*{Acknowledgments}
We thank A.~Mazeliauskas, N.~Mueller, S.~Schlichting, and R.~Venugopalan for valuable discussions.
This work is part of and supported by the Deutsche Forschungsgemeinschaft (DFG) Collaborative Research Centre ``SFB~1225~(ISOQUANT)".
We acknowledge support by the state of Baden-W\"{u}rttemberg through bwHPC. 
Part of this work was performed on the computational resource ForHLR II, funded by the Ministry of Science, Research and the Arts Baden-W\"{u}rttemberg and DFG.

\appendix
\section{Removal of doublers} \label{sec:doub}
In this appendix, we show results that justify the method to remove doublers introduced in Sec.~\ref{subsec:initial}. In that method, the doublers are removed in the initial condition, Eq.~\eqref{stoPsi}, and the evolution equation is not modified. We will refer to this procedure as initial condition (IC) removal method, and compare it with the Wilson fermion method.

In the Wilson fermion method, fermion doublers are suppressed by a Wilson term $W \psi$ in the Dirac equation,
\begin{equation}
\left[ i\slashchar{D} -m +W \right] \psi = 0 \, .
\end{equation}
In contrast to conventional Euclidean lattice simulations,
where such a Wilson term is added to suppress all temporal and spatial doubler modes, in real-time calculations typically only a Wilson term for the suppression of spatial doublers is added while temporal doublers are suppressed in the initial conditions~\cite{Saffin:2011kc}. For longitudinally expanding systems this has to be suitably generalized. A possible form of the (improved) spatial Wilson term in the longitudinally expanding geometry is
\begin{align}
W\psi (x) &= \frac{r_\perp}{2a_\perp} \sum_{i=1}^2 \left\{ 
c_1 \left[ U_i(x) \psi (x+\hat{i}) -2\psi(x) +U_i^\dagger (x-\hat{i}) \psi (x-\hat{i} )\right] \right. \notag \\
&\hspace{10pt} \left. 
+2c_2\left[ U_i(x) U_i(x+\hat{i}) \psi (x+2\hat{i}) -2\psi(x) +U_i^\dagger (x-\hat{i}) U_i^\dagger (x-2\hat{i}) \psi (x-2\hat{i} )\right] \right\} \notag \\
&\hspace{10pt}
+\frac{r_\eta}{2Ta_\eta} \left\{
c_1 \left[ U_\eta(x) \psi (x+\hat{\eta}) -2\psi(x) +U_\eta^\dagger (x-\hat{\eta}) \psi (x-\hat{\eta} )\right] \right. \notag \\
&\hspace{10pt} \left.
+2c_2 \left[ U_\eta(x) U_\eta(x+\hat{\eta}) \psi (x+2\hat{\eta}) -2\psi(x) +U_\eta^\dagger (x-\hat{\eta}) U_\eta^\dagger (x-2\hat{\eta}) \psi (x-2\hat{\eta} )\right] \right\} \, ,
\label{Wilson}
\end{align}
where $r_\perp$ and $r_\eta$ are arbitrary parameters, and $T$ is a quantity that has the dimension of time.
One natural choice for the latter quantity is $T=\tau$. In this case, the Wilson term amounts to a time-dependent mass term. One of the consequences is that the definition of a fermion quasi-particle distribution becomes more problematic.
Another possibility is to identify $T$ with a fixed time. 
In order to explore the impact of such longitudinal Wilson terms in the longitudinally expanding geometry, we implement two cases for comparison: 
\begin{itemize}
\item[(i)] Only the transverse Wilson term, $r_\perp=1$ and $r_\eta=0$. 
\item[(ii)] The full spatial Wilson term with $r_\perp=1$,  $r_\eta=1$ and $T=\tau_0$. 
\end{itemize}
In the Wilson fermion method, the fermion fields are initialized according to Eq.~\eqref{stoPsi0} with the dispersion relation modified by the Wilson term. 

\begin{figure}[tb]
 \begin{center}
   \includegraphics[clip,width=8cm]{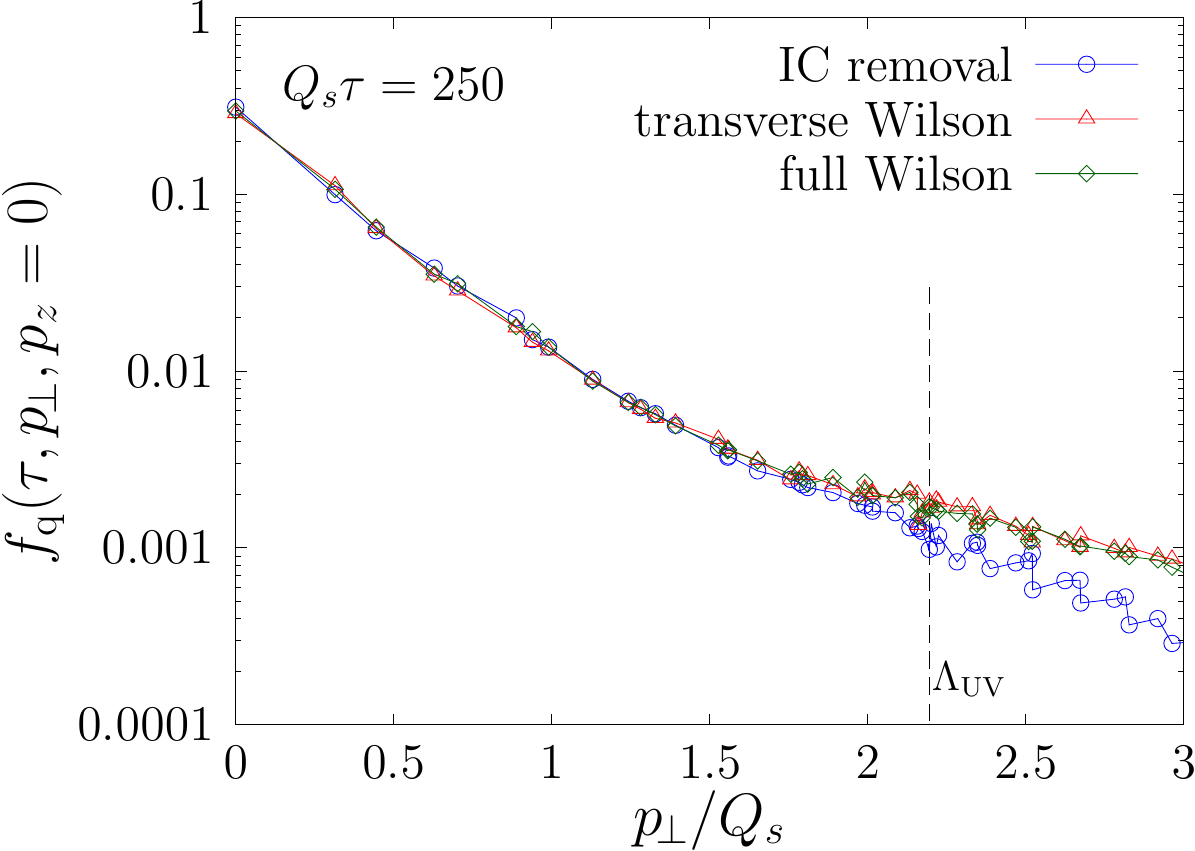}
 \end{center}
 \vspace{-20pt}
 \caption{Comparison of the three methods for the transverse distribution function at $Q_s \tau=250$.}
\label{fig:doubdep_distript}
\end{figure}

In Fig.~\ref{fig:doubdep_distript}, the three different methods are compared for the transverse momentum distribution of quarks in the physical momentum region at the time $Q_s \tau=250$. These computations are done on a lattice of size $N_\perp =32$, $N_\eta=256$, $Q_s a_\perp =0.625$, and $a_\eta=1.95 \times 10^{-3}$. 
The three different procedures agree well for $\pperp <\Lambda_\text{UV}$, where $\Lambda_\text{UV}$ is the ultraviolet (UV) cutoff in one transverse direction. Since we plot the distribution as a function of $\pperp =\sqrt{p_x^2 +p_y^2}$, there are data points in the region $\pperp >\Lambda_\text{UV}$. In that region, the IC removal method shows a deviation from the Wilson fermion method. However, this region is affected by the cutoff which would become irrelevant approaching the continuum limit.\footnote{In the figures shown in the main sections, this region is not plotted.} 

\begin{figure}[tb]
 \begin{center}
   \includegraphics[clip,width=15cm]{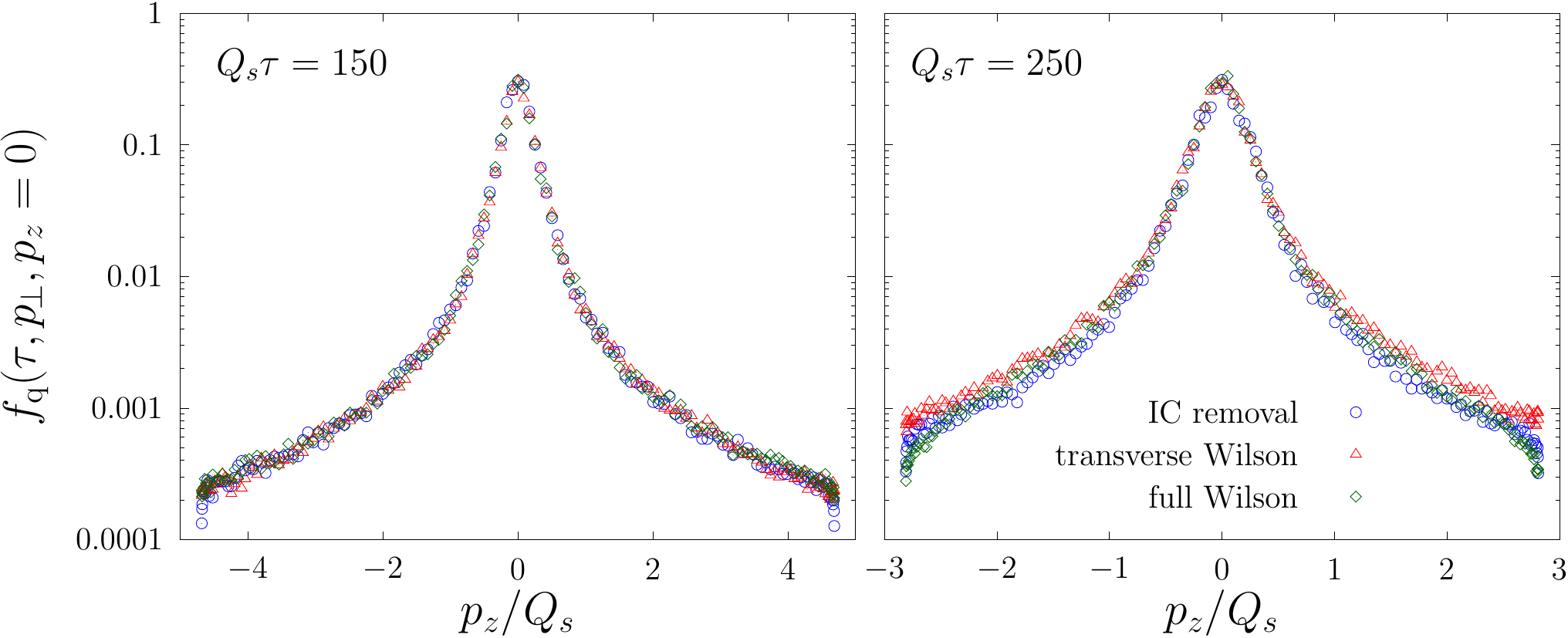}
 \end{center}
 \vspace{-20pt}
 \caption{Comparison of the three methods for the longitudinal distribution function at $Q_s \tau=150$ and $Q_s \tau=250$.}
\label{fig:doubdep_distripz}
\end{figure}

The same comparison is made for the longitudinal distribution in Fig.~\ref{fig:doubdep_distripz}. Since the longitudinal distribution is more affected by the expansion of the system, it is plotted for two different times. At $Q_s \tau=150$, the three results lie on top of each other, indicating that the different ways to remove doublers do not affect the physical region. At a later time, $Q_s \tau=250$, the results with the IC removal method and the full Wilson term still show a good agreement. However, the result with the transverse Wilson term exhibits deviations at large momentum regions. This is natural, because the longitudinal doublers are not suppressed in this method and they can interact with the physical modes.  

\begin{figure}[tb]
\begin{tabular}{ccc}
 \begin{minipage}{0.48\hsize}
 \begin{center}
   \includegraphics[clip,width=7.5cm]{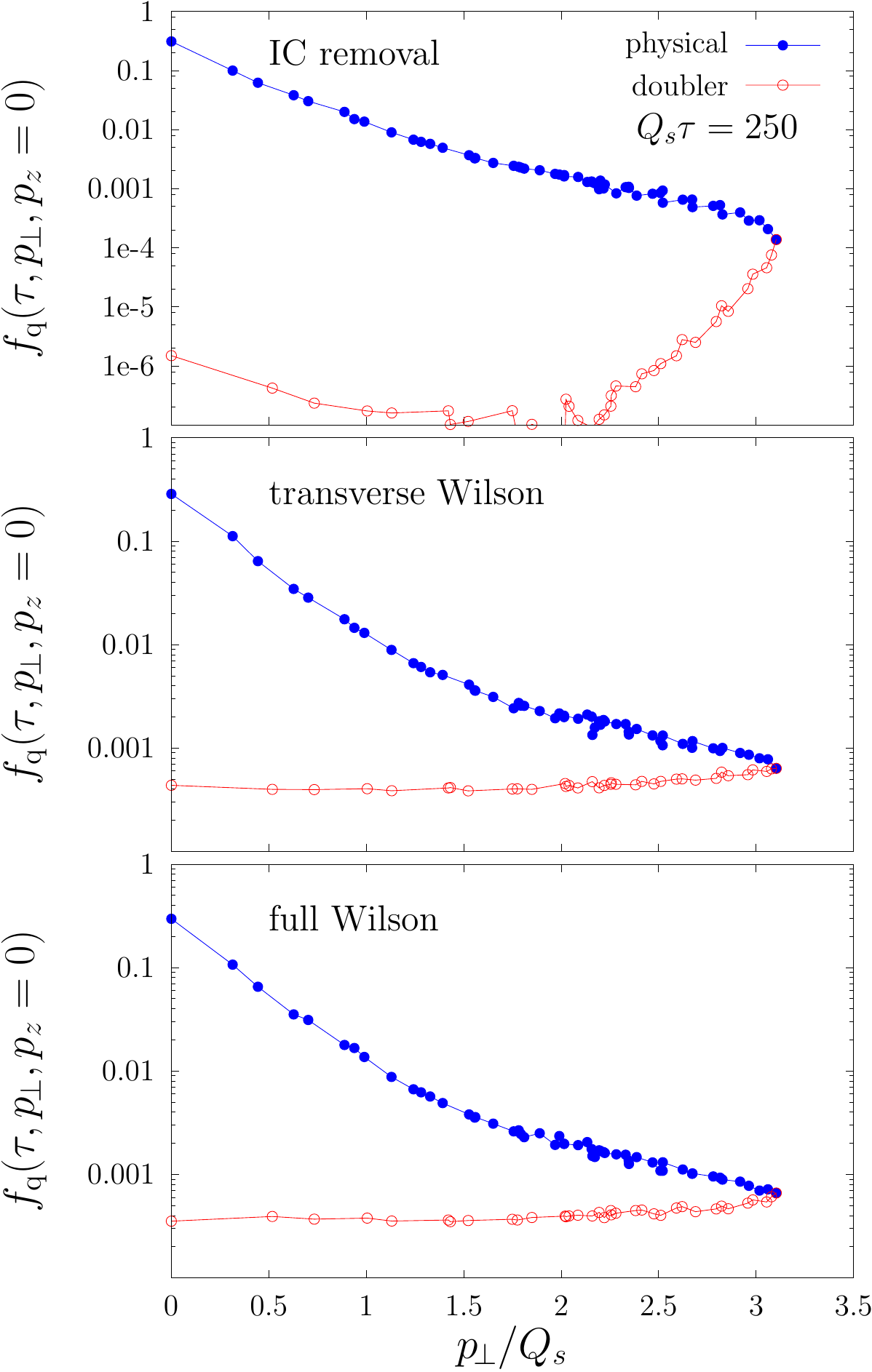}
 \end{center}
 \vspace{-20pt}
 \caption{Comparison of the transverse distributions for physical modes and doubler modes computed by three methods.}
 \label{fig:doubdep_distript_d}
 \end{minipage} & \ &
 \begin{minipage}{0.48\hsize}
 \begin{center}
   \includegraphics[clip,width=7.5cm]{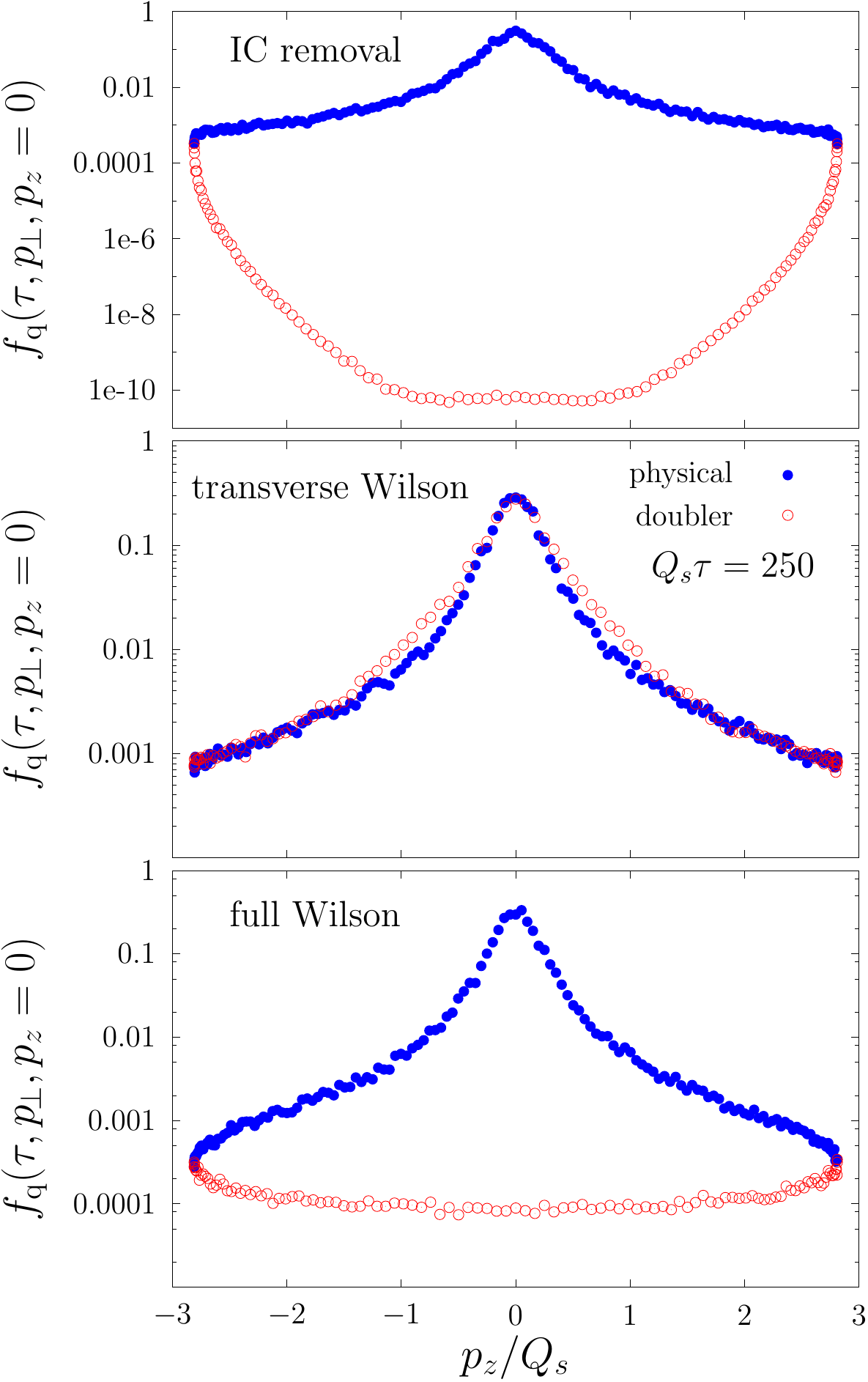}
 \end{center}
 \vspace{-20pt}
 \caption{Comparison of the longitudinal distributions for physical modes and doubler modes computed by three methods.}
 \label{fig:doubdep_distripz_d}
 \end{minipage}
 \end{tabular}
\end{figure}

To have a closer look at how doubler modes are suppressed, we directly compare the distributions for the physical modes and for the doubler modes in each method. As shown in Fig.~\ref{fig:doubdep_distript_d}, the transverse doubler modes are suppressed by all the methods. For the longitudinal distributions plotted in Fig.~\ref{fig:doubdep_distripz_d}, we can see that the IC removal method and the full Wilson term successfully suppress doubler modes. In the result with the transverse Wilson term, the longitudinal doublers are not suppressed as expected. Since the longitudinal UV cutoff is decreasing in time, the longitudinal doubler modes starts to interact with the physical modes at later times resulting in the deviation seen in Fig.~\ref{fig:doubdep_distripz}. 

To summarize, we have demonstrated that both of the IC removal method and the Wilson fermion method with a choice of $T=\tau_0$ can suppress doubler modes successfully without affecting the physical modes for the times of interest in this study. In the main part of this paper, we show numerical results computed in the IC removal method as its numerical cost is lower than for the Wilson fermion method. 

Two final remarks are in order.
To address the physics related with the chiral anomaly, one cannot use the IC removal method, since this method amounts to introducing the UV cutoff to canonical momentum and therefore cannot describe the chiral anomaly \cite{Tanji:2016dka}. 
We expect that the chiral anomaly in the expanding geometry can be described by using the full Wilson term. However, with a choice of $T=\tau_0$, there is a limitation associated with the separation of time scales. The longitudinal Wilson term is time-independent for a fixed $T$, while the longitudinal momentum scales decrease in time as $1/\tau$. Therefore, the longitudinal Wilson term would start affecting the physical modes at some later times.
We will leave the investigation of the chiral anomaly on the longitudinally expanding lattice to future work.

\section{Cutoff dependence} \label{sec:cutoff}
In this appendix, we discuss lattice cutoff dependencies of physical quantities.
The UV momentum cutoff $\Lambda_\text{UV}$ and the smallest nonzero momentum $\Lambda_\text{IR}$, which plays a role of the infrared (IR) cutoff, are related with the lattice spacing $a$ and the system size $L$, respectively, as
\begin{gather}
\Lambda_\text{UV} \simeq 1.37/a \, , \hspace{10pt}
\Lambda_\text{IR} \simeq 2\pi/L \, .
\end{gather}

\begin{figure}[tb]
 \begin{center}
  \includegraphics[clip,width=8cm]{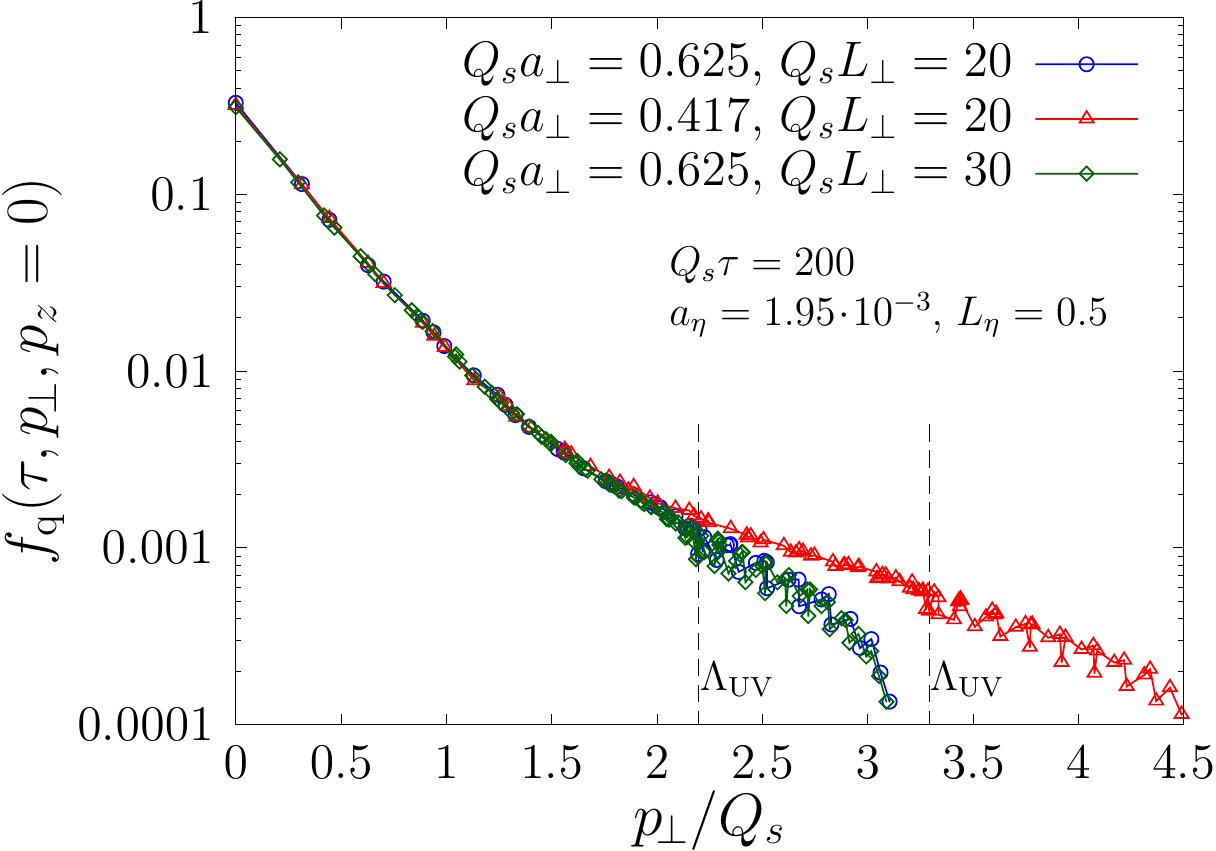} 
  \vspace{-10pt}
  \caption{Transverse momentum distributions at $Q_s \tau=200$. 
  Three different transverse lattice parameters are compared.}
  \label{fig:lattdep_pt}
 \end{center}
\end{figure}

\begin{figure}[tb]
 \begin{center}
  \includegraphics[clip,width=8cm]{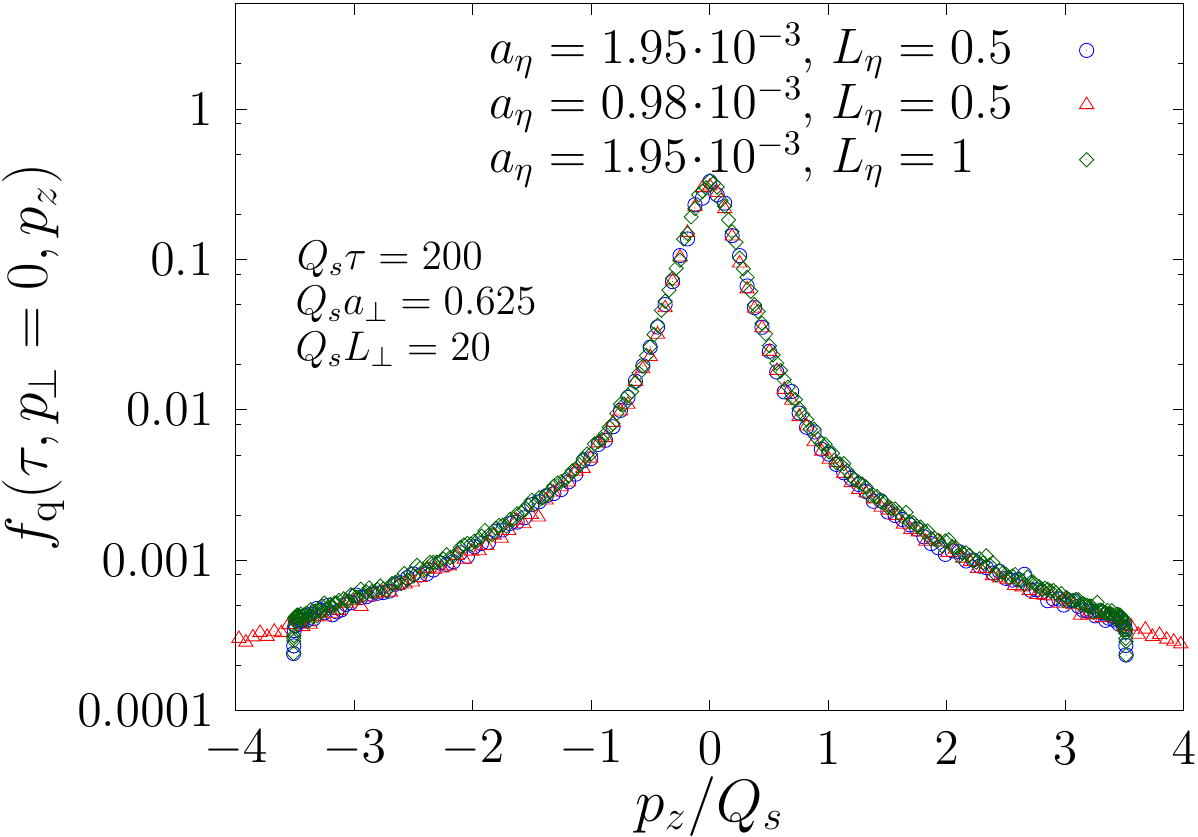} 
  \vspace{-10pt}
  \caption{Longitudinal momentum distributions at $Q_s \tau=200$. 
  Three different longitudinal lattice parameters are compared.}
  \label{fig:lattdep_pz}
 \end{center}
\end{figure}

Figure \ref{fig:lattdep_pt} shows the transverse momentum distributions at $p_z=0$ and $Q_s \tau =200$ for three different sets of the transverse lattice parameters. Two vertical dashed lines indicate the UV cutoff in one transverse direction that  corresponds to the lattice spacing $Q_s a_\perp =0.625$ and $Q_s a_\perp =0.417$, respectively. As discussed in Fig.~\ref{fig:doubdep_distript}, there is a region where $\pperp =\sqrt{p_x^2+p_y^2}$ is larger than $\Lambda_\text{UV}$. 
We can confirm that the distribution function for $p_\perp <\Lambda_\text{UV}$ is rather insensitive to variations of either the UV cutoff or the IR cutoff. 
Similarly, the longitudinal momentum distributions for three different sets of the longitudinal lattice parameters are plotted in Fig.~\ref{fig:lattdep_pz}. Also, in this case, both of the UV and the IR cutoffs do practically not affect the shape of the distribution.

\begin{figure}[tb]
 \begin{tabular}{cc}
 \begin{minipage}{0.5\hsize}
  \begin{center}
   \includegraphics[clip,width=7.8cm]{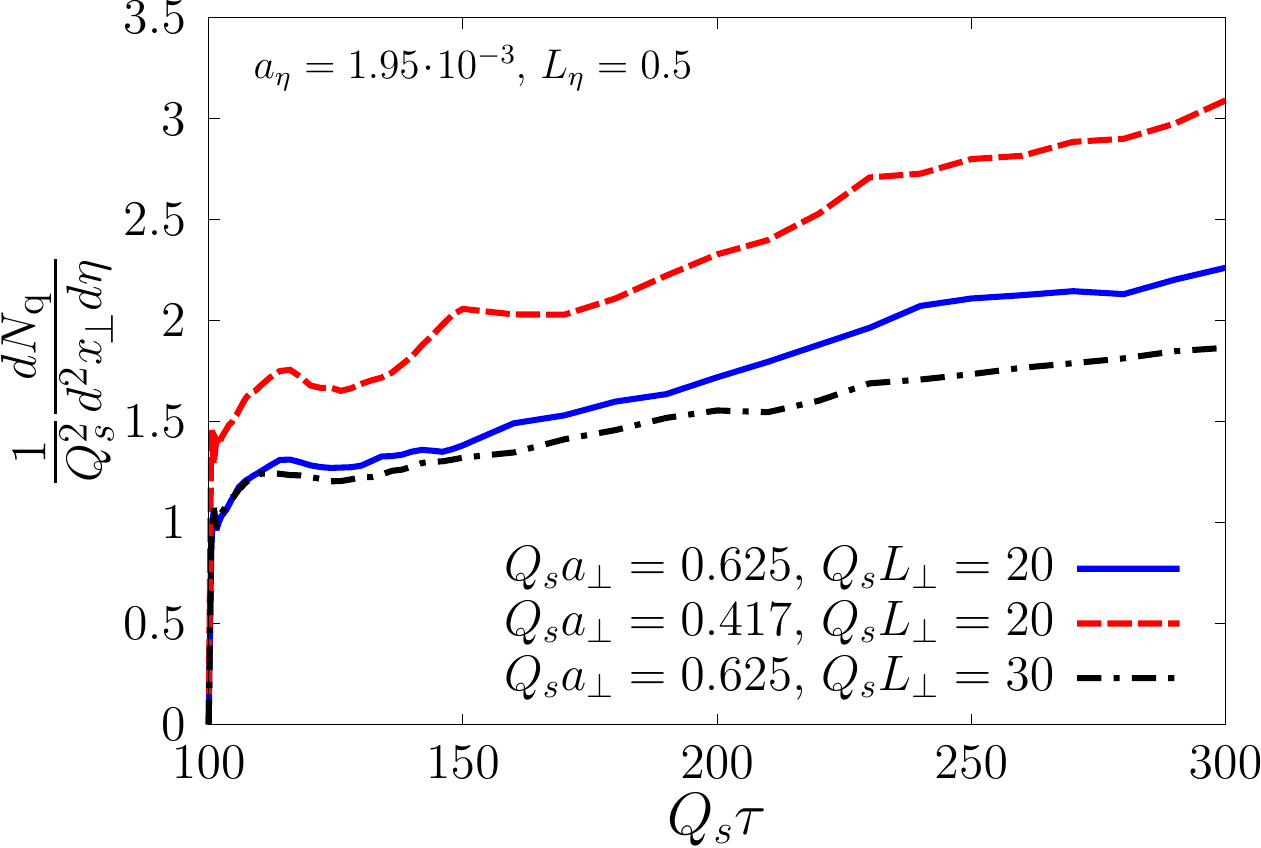}
  \end{center}
 \end{minipage} &
 \begin{minipage}{0.5\hsize}
  \begin{center}
   \includegraphics[clip,width=7.8cm]{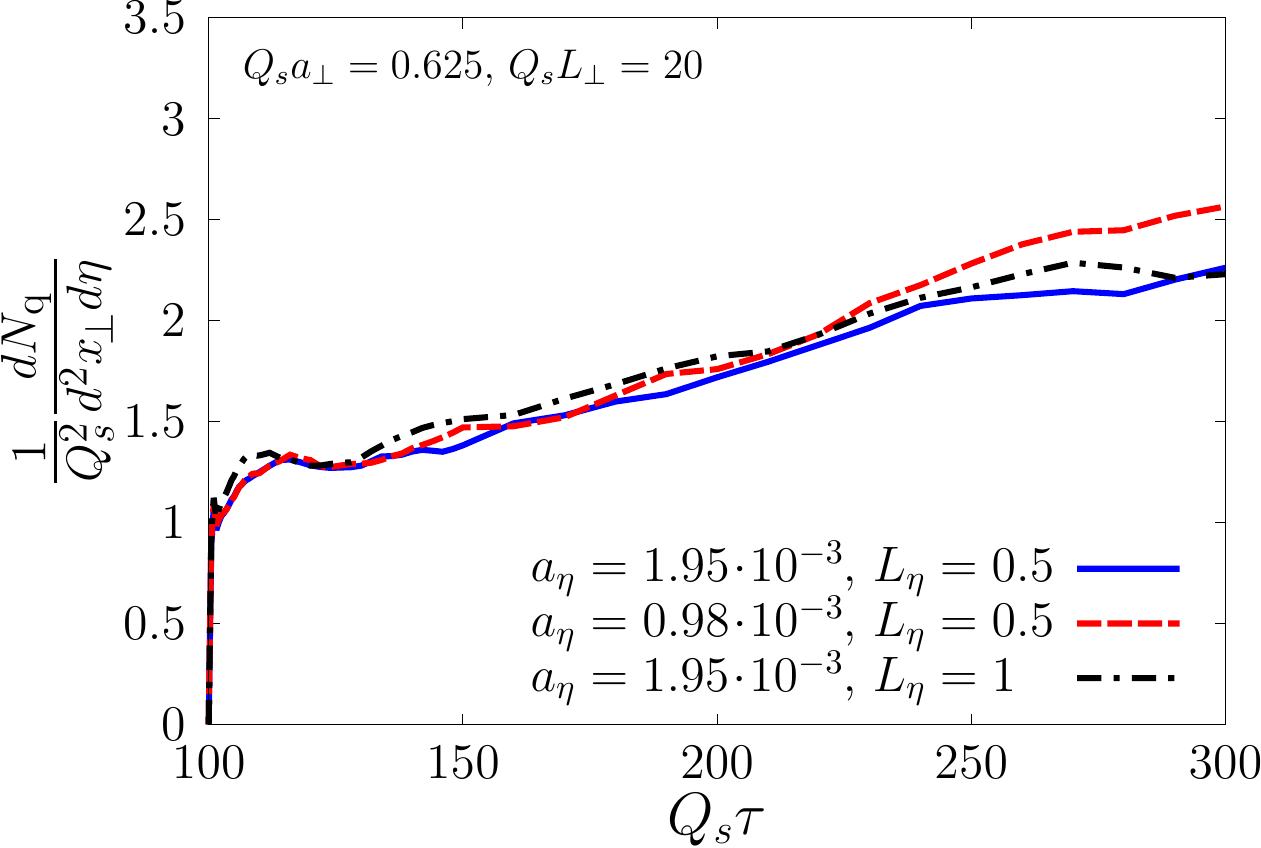}
  \end{center}
 \end{minipage} 
 \end{tabular}
\caption{Time evolution of the quark number density for different lattice parameters.
Left: Dependence on transverse lattice parameters. Right: Dependence on the longitudinal lattice parameters.}
\label{fig:lattdep_numq}
\end{figure}

Next, we check the cutoff dependence of the total quark number density. 
Figure \ref{fig:lattdep_numq} shows the time evolution of the quark number density for different transverse lattice parameters (left) and for different longitudinal lattice parameters (right). 
While the dependence on the longitudinal lattice parameters is relatively minor, the dependence on the transverse lattice parameters is notable. This is because the contribution from the high $\pperp$ tail that is contaminated by the UV cutoff 
is still large for the lattice parameters we employ. As shown in Fig.~\ref{fig:lattdep_pt}, the shape of the distribution in the region $\pperp >\Lambda_\text{UV}$ depends on the UV cutoff. Since the phase-space factor enhances the contribution from the high $\pperp$ tail, the integrated particle number exhibits strong dependence on the UV cutoff. 
We expect that the dependence on the transverse UV cutoff would become milder if the cutoff is sufficiently large. 
Indeed, we have obtained smaller dependence on the longitudinal UV cutoff as we have larger longitudinal UV cutoff than transverse in our computations.
Furthermore, as discussed in Fig.~\ref{fig:Debye1}, the system size needs to be sufficiently large in order to resolve the IR sector, and hence to correctly describe the later-time kinetic processes.


\end{document}